\documentclass{aa}  

\usepackage{graphicx}
\usepackage{txfonts}
\usepackage{lipsum}
\usepackage{subcaption}         % necessary for continued figures, example in section 3
\usepackage{lscape}             % to rotate a single page table, example in appendix.
\usepackage{placeins}           % useful with \FloatBarrier, to keep 
 \usepackage{longtable}                                % onecolumn floats from drifting to the next section

\usepackage{amsmath}
\begin{document}

   \title{JWST MIRI color classification of mid-infrared selected galaxies}

   \subtitle{MIRI color classification toward cosmic noon}

   \author{E. Kilerci
          \inst{1}\fnmsep\thanks{Corresponding author}
          \and
          T. Goto\inst{2}
          \and
          M. A. Malkan\inst{3}
          \and
          S. J. Kim\inst{2}
          \and
          C.-T. Ling \inst{4,5}
          \and
          C. K.-W. Wu \inst{2}
          \and
          T. Hashimoto \inst{6}
          \and
          S. C.-C. Ho \inst{7}
          \and
         Amos Y.-A. Chen \inst{2}
          \and
          E. G\"{o}\u{g}\"{u}\c{s} \inst{8}
          }

   \institute{Department of Astronomy and Space Sciences, Science Faculty, \.{I}stanbul University, Beyaz{\i}t 34119, \.Istanbul, T\"urkiye\\
              \email{ecekilerci@istanbul.edu.tr}
         \and
             Institute of Astronomy, National Tsing Hua University, 101, Section 2. Kuang-Fu Road, Hsinchu, 30013, Taiwan (R.O.C.) \\
        \and
             Department of Physics and Astronomy, University of California Los Angeles, 430 Portola Plaza, Los Angeles, CA 90095, USA\\
             %\email{}
         \and
            Department of Astronomical Science, SOKENDAI (The Graduate University for Advanced Studies), 2-21-1 Osawa, Mitaka, Tokyo, 181-8588, Japan\\
         \and 
             National Astronomical Observatory of Japan, 2-21-1 Osawa, Mitaka, Tokyo 181-8588, Japan\\
        \and
             Department of Physics, National Chung Hsing University, No. 145, Xingda Rd., South Dist., Taichung, 40227, Taiwan (R.O.C.)\\
             %\email{}
        \and
             Research School of Astronomy and Astrophysics, The Australian National University, Canberra, ACT 2611, Australia\\
             %\email{} 
         \and
             Sabanc{\i} University, Faculty of Engineering and Natural Sciences, 34956, Istanbul, T\"urkiye\\
             %\email{}
             }

   \date{Received XXX; accepted YYY}

% \abstract{}{}{}{}{} 
% 5 {} token are mandatory
  \abstract
  % context heading (optional)
  % {} leave it empty if necessary  
   {We investigated the \textit{James Webb} Space Telescope photometric color classification of mid-infrared (MIR) selected galaxies at high redshifts, toward cosmic noon.}
  % aims heading (mandatory)
   {The aim of the present work is to obtain a $z$-dependent mid-infrared (MIR) photometric galaxy classification tool based on broad spectral emission and absorption lines using the JWST Mid-Infrared Instrument (MIRI) and its broadband filters.} 
  % methods heading (mandatory)
   {We used the largest \textit{Spitzer} MIR spectral database to obtain synthetic photometry in the JWST/MIRI filters. We formed MIRI filter combinations to trace the strong polycyclic aromatic hydrocarbon (PAH) emission features and the 9.7\,$\mu$m silicate feature in seven redshift windows from $z=0.25-2.10$. }
  % results heading (mandatory)
   {We present $z$-dependent MIRI color--color plots that separate active galactic nuclei (AGN), star-forming galaxies (SFGs), and silicate absorption-dominated galaxies up to $z \sim 2$.  We applied the photometric MIR colors to the largest ($\sim34$\,arcmin$^2$) MIRI survey called the Systematic Mid-infrared Instrument Legacy Extragalactic Survey (SMILES), to identify AGN, SFGs, and Si-absorption dominated galaxies out to substantial redshifts. Our JWST/MIRI SFGs sample includes galaxies with total IR luminosities of $10^{9.2}$ $\sim$ $10^{11.9}$ $L_{\odot}$ at $0.9\leq z < 1.57$. The majority of them are consistent with the $z \sim 1$ main sequence. 
   We also identified the first examples of $z \sim 1$ galaxies with deep silicate absorption.}
  % conclusions heading (optional), leave it empty if necessary 
   {}


   \maketitle
%
%-------------------------------------------------------------------

\section{Introduction} \label{sec:intro}
The mid-infrared (MIR) spectra of galaxies are essential probes into the nature of different populations, such as active galactic nuclei (AGN) and star-forming galaxies (SFGs), even when their optical emission is largely obscured by dust. Distinguishing galaxy populations at higher-$z$, toward cosmic noon, is important for tracking the evolution of cosmic star formation \citep[(CSFH) e.g.,][]{Magnelli2013} and the black hole accretion history \citep[(BHAH) e.g.,][]{Yang2023agn} across cosmic time. The strongest spectral features in the MIR wavelengths are the polycyclic aromatic hydrocarbon (PAH) emission features at 3.3\,$\mu$m, 6.2\,$\mu$m, 7.7\,$\mu$m, 11.2\,$\mu$m, 12.7\,$\mu$m, (hereafter PAH33, PAH62, PAH77, PAH112, PAH127) and the silicate (Si) emission and absorption features at 9.7\,$\mu$m (hereafter Si97) and 18\,$\mu$m. While the PAH emission lines are tracers of star formation activity \citep[see, e.g.,][]{Peeters2004,Farrah2007,Calzetti2007}, the silicate absorption features are generated by a bright continuum from AGN or starbursts, passing through a large amount of silicate dust producing Si$-$O stretching mode absorption. Therefore, deep silicate absorption features are uniquely capable of revealing the most heavily obscured AGN \citep[see, e.g.,][]{Spoon2001,Spoon2006,Levenson2007} or compact starbursts \citep[see, e.g.,][]{Armus2004,Spoon2007}. 

In the pre-JWST era, the Infrared Space Observatory \citep[\textit{ISO},][]{Kessler1996} and \textit{Spitzer} space telescope's \citep{Werner2004} Infrared Spectrograph \citep[IRS,][]{Houck2004} provided MIR spectra of mostly low-redshift galaxies. \citet[][]{Spoon2007} examined the IRS and \textit{ISO} spectra of AGN, starburst galaxies, infrared galaxies (IRGs), luminous infrared galaxies (LIRGs), and ultraluminous infrared galaxies (ULIRGs), and divided them into nine groups (ranging between continuum-, AGN-, PAH-, or absorption- dominated MIR spectra) according to the equivalent width (EQW) of the 6.2\,$\mu$m emission line and the strength of the 9.7\,$\mu$m silicate feature \citep[$S_{\rm{sil}}$, the negative of the apparent Si optical depth defined as][Eq. (1)]{Spoon2007}. An extended study of more than 3000  \textit{Spitzer} IRS spectra \citep[][]{Spoon2022} show that by using the PAH62 emission line EQW (hereafter PAH62 EQW) and the $S_{\rm{sil}}$ galaxies can be sorted into three main MIR spectral classes: (1) star-forming galaxies with strong PAH emission and weak silicate features \citep[referred as `1C' galaxies in][]{Spoon2022}; (2) strongly AGN-dominated galaxies with no or little PAH emission and weak silicate emission or absorption \citep[referred as `1A' galaxies in][]{Spoon2022};  (3) Si absorption-dominated galaxies with weak or undetectable PAH emission, whose power source can be an ultracompact nuclear starburst or AGN \citep[referred as `3A' galaxies in][]{Spoon2022}. To cover the large range of MIR spectra of galaxies, five additional intermediate spectral classes are introduced in the diagnostic diagram of \citet{Spoon2022} (their Fig. 14): (i) AGN with moderate PAH features as in normal star-forming galaxies \citep[referred as `1B' galaxies in][]{Spoon2022}; ii) dusty star-forming galaxies with moderate PAH features and $S_{\rm{sil}}$ \citep[referred as `2B' galaxies in][]{Spoon2022}; (iii) dusty star-forming galaxies with strong PAH features \citep[referred as `2C' galaxies in][]{Spoon2022}; (iv) dusty AGN with moderate $S_{\rm{sil}}$ \citep[referred as `2A' galaxies in][]{Spoon2022}; (v) Si absorption-dominated SFGs galaxies with PAH emission \citep[referred as `3B' galaxies in][]{Spoon2022}. 
 Dusty systems such as ULIRGs (with IR luminosities ($L_{\rm IR}$) 10$^{12}L_{\sun} <$ $L_{\rm IR}<$ 10$^{13}$ $L_{\sun}$) and LIRGs (10$^{11}L_{\sun} <$L$_{IR}<$ 10$^{12}$ L$_{\sun}$) \citep[e.g.,][]{Sanders1988b,Veilleux2002} are mainly in classes 2A–3B \citep[e.g.,][]{Armus2004,Spoon2007}.  Based on the galaxy evolution scenarios of U/LIRGs \citep[e.g.,][]{Hopkins2008b}, the MIR spectral diagnostic diagram may be interpreted as an evolutionary sequence starting from 1C or 1B mergers and reaching the 1A pure AGN phase after the dust-covered phase of interacting galaxies in 2A--3B \citep[][]{Spoon2022}.

Since large spectroscopic surveys in the MIR are not practical, 
the MIR photometric color selection of different galaxies, such as AGN, obscured AGN, compact obscured nuclei (CONs), star-forming galaxies (SFGs), and AGN+star-forming  composites \citep[e.g.,][]{Stern2012,Mateos2013,kirkpatrick2013,Kirkpatrick2017b, Rovilos2014,Ichikawa2019,GarciaBernete2022,Kirkpatrick2023}, is a powerful method to select galaxies for spectroscopic follow-ups and perform statistical analysis of large samples.  
The Mid-Infrared Instrument \citep[(MIRI),][]{Rieke2015,Wright2023} of JWST provides continuous photometric coverage with its narrow and broad bands in the range $\sim$5--25 $\mu$m, with a uniquely high sensitivity, down to 1\,$\mu$Jy \citep[e.g.,][]{Ling2022, Wu2023, Yang2023miri, Kirkpatrick2023}. This unprecedented sensitivity allows us, for the first time, to identify fainter galaxies with $L_{\rm IR}=10^{9}-10^{10} L_{\sun}$ at $z = 1 - 2$ \citep{Kirkpatrick2023}. MIRI performs imaging with nine filter bandpasses, F560W, F770W, $F1000W$, $F1280W$, $F1500W$, $F1800W$, $F2100W$, and $F2550W$,  centered at 5.6-, 7.7-, 10.0-, 11.3-, 12.8-, 15.0-, 18.0-, 21.0-, and 25.5\,$\mu$m; respectively.
Before the launch of JWST, \citet{Kirkpatrick2017b} used IR SED templates of high-redshift dust-rich galaxies with strong PAH emission and presented a redshift-dependent MIRI color selection covering the $z=1-2$ range to identify AGN, SFGs, and composite mixtures of both. Their color selection traces the 6.2 and 7.7\,$\mu$m PAH features and the 3–5\,$\mu$m stellar minimum at $z\sim$1, $z\sim$1.5, and $z\sim$2. 
\citet{Kirkpatrick2023} present an updated color selection (which does not require redshift information) sensitive to PAH62 and PAH77 that separates AGN, SFGs, and MIR weak galaxies based on the MIRI survey obtained for The Cosmic Evolution Early Release Science (CEERS) program in the Extended Groth Strip (EGS). 
\citet[][]{GarciaBernete2022} presents MIRI color diagrams to select the most obscured galaxy nuclei in CONs based on SED model fitting results of local U/LIRGs selected from \textit{Spitzer} IRS spectra sample of \citet[][]{Spoon2022}.
\citet{Lin2024} introduced a color--color diagram to select particularly bright sources at 15\,$\mu$m that are likely PAH luminous galaxies at $z\sim$1 in the CEERS MIRI survey. 
MIR selection methods have long been used to identify AGN \citep[e.g.,][]{Stern2012,Assef2013}. More recently, \citet{Bornancini2020} analyzed the host galaxy properties of obscured and unobscured AGN selected in the MIR, providing complementary insights into the nature of these populations. These studies provide a broader context for our approach, which extends MIR diagnostics into the JWST/MIRI regime with redshift-dependent color cuts.

With the new sensitivity and continuous photometric coverage of MIRI, the MIR spectral features of galaxies can be traced out to cosmic noon ($z \sim 1-2.0$) \citep[e.g.,][]{Yang2023miri, Kirkpatrick2023}. 
The aim of this study is to develop and apply a $z$-dependent color--color diagnostic for the three main MIR spectral classes defined by PAH62 EQW and $S_{\rm{sil}}$ \citep[][]{Spoon2022}. We focus on identifying SFGs, AGN-dominated, and Si absorption-dominated galaxies by tracing PAH62, PAH77, PAH112, PAH127, and Si 9.7\,$\mu$m  through MIRI bands over the $0.25\le z \le 2.10$ range, where the available \textit{Spitzer} IR spectra are very sparse (e.g., there are just a few \textit{Spitzer} spectra for Si absorption-dominated galaxies in the 3A and 3B classes at $z\ge1$). We used the largest \textit{Spitzer} spectral database, The Infrared Database of Extragalactic Observables \citep[IDEOS sample;][]{Hernan-Caballero2016, Spoon2022} for our investigation. 

The structure of this paper is as follows. 
We present the IDEOS sample in \S \ref{S:sample}. 
In \S \ref{S:colcol} we introduce our color selection criteria for SFGs, AGN-dominated galaxies, and Si absorption-dominated galaxies at different $z$ values.
We present the SMILES JWST/MIRI photometric survey, to which we apply our color diagnostics, in Sect. \ref{S:smiles}.
In \S \ref{S:colsmiles} we describe the MIRI galaxy classification in the SMILES sample.
The main results and discussion of our analysis are presented in \S \ref{S:results}. 
Our main conclusions are summarized in \S \ref{S:conc}.
In this work we adopt a flat $\Lambda$ cold dark matter cosmology with 
$H_0=72$\,km\,s$^{-1}$\,Mpc$^{-1}$, $\Omega_\Lambda = 0.7$, and $\Omega_{\rm m}=0.3$ \citep{Spergel2003}. 

\section{Sample and analysis}\label{S:sampleandanalysis}

\subsection{\textit{Spitzer} IDEOS sample}\label{S:sample}

The IDEOS database \citep[][]{Hernan-Caballero2016, Spoon2022} provides homogeneously analyzed archival \textit{Spitzer} IRS spectra of 3335 galaxies with $S_{\rm{sil}}$, PAH line fluxes and equivalent widths, rest-frame continuum flux densities, synthetic MIRI photometry measurements, and MIR classification. To investigate redshift-dependent MIRI colors of different types of MIR galaxies we used \textit{Spitzer} IRS spectra from the IDEOS sample. To represent galaxies in different MIR classes we selected spectra with signal-to-noise ratio (S/N) $\geq 5$ at least in two continuum regions at rest frame 6.6\,$\mu$m and 9\,$\mu$m or 6.6\,$\mu$m and 13.25\,$\mu$m or 9\,$\mu$m and 11.25\,$\mu$m.  This selection yielded large numbers of spectra in each MIR class: 805 1A, 307 1B, 494 1C, 50 2A, 134 2B, 72 2C, 19 3A, and 31 3B. In total, our IDEOS sample included 1912 spectra.
We convolved these spectra with MIRI filters following the synthetic photometric flux density recipe given by \citet{Spoon2022}, namely, we measured the photon-weighted mean flux density over the bandpass of the filter as given by their equation 7.

\subsection{Color analysis}\label{S:colcol}
As noted by \citet{Spoon2022}, their MIR spectral classification is successful in distinguishing three main classes: pure SFGs (1C), AGN-dominated galaxies (1A), and Si absorption-dominated galaxies (3A). Since we rely on their MIR classification, here we search for colors to separate especially the pure SFGs, AGN-dominated galaxies, and Si absorption-dominated galaxies between $0.1 \leq z \leq 2$. First, we expanded the MIR spectral classification based on PAH62 to other PAH lines since the PAH62 EQW of our IDEOS sample strongly correlates with  PAH77 EQW, PAH112 EQW, PAH127 EQW as shown in Appendix \ref{S:appendix}. We examined various color--color combinations to trace PAH emission features, Si absorption feature, and hot-dust-dominated continuum. We determined the $z$ ranges at which the Si and PAH features can be probed in MIRI bands. The 9.7 $\mu$m Si absorption feature shifts to 9.7(1+$z$); at $z=$ 0.25--0.30, 0.58--0.65, 0.9--0.92, 1.0--1.1, and 1.5--1.6 it enters the 12.8\,$\mu$m, 15.0\,$\mu$m, 18.0\,$\mu$m, 21.0\,$\mu$m and  25.5\,$\mu$m imaging bands, respectively. Therefore, it can be traced by MIRI up to $z \sim 1.6$. 
The examined MIRI color--color diagrams combining Si97 and PAH62, PAH77, PAH112, PAH127 colors at different redshift intervals are shown in Fig. \ref{fig:fig1}.

\subsubsection{Classification of each MIR class}\label{S:eachclass}

\begin{figure*}
\begin{center}$
\begin{array}{lll}
\includegraphics[scale=0.5]{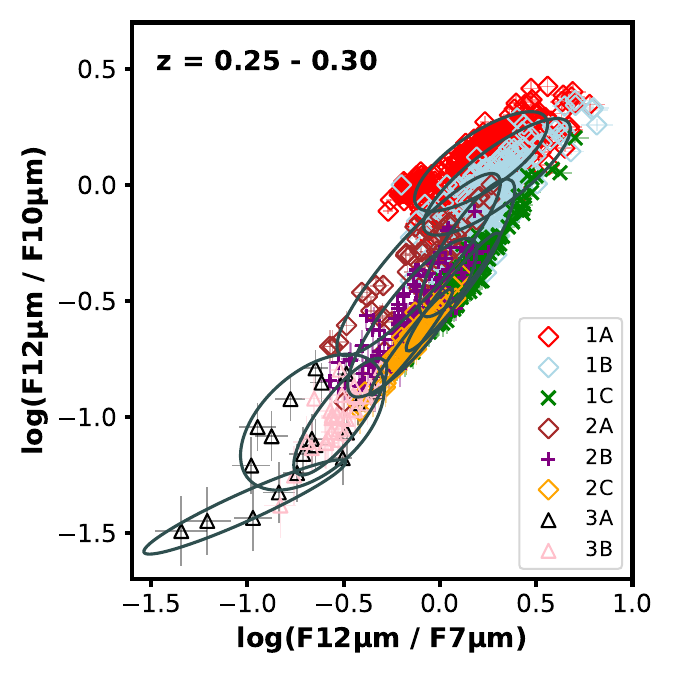}&
\includegraphics[scale=0.5]{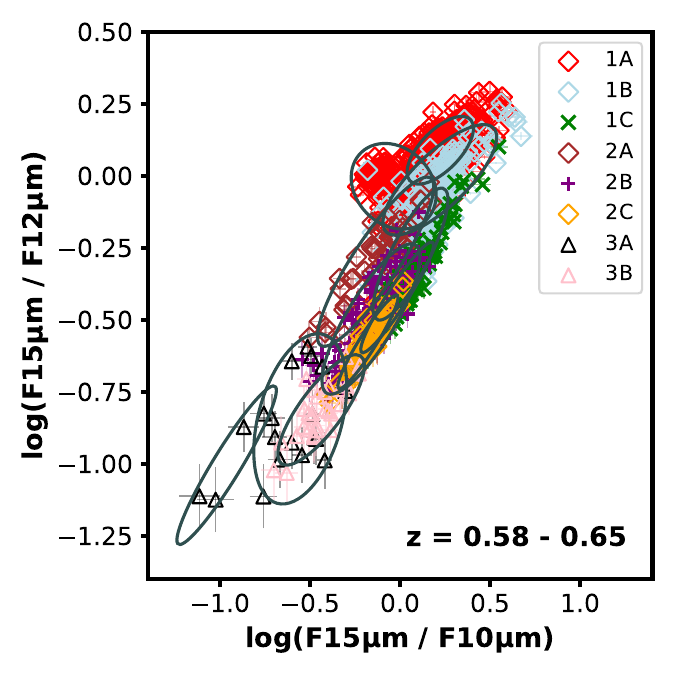}&
\includegraphics[scale=0.5]{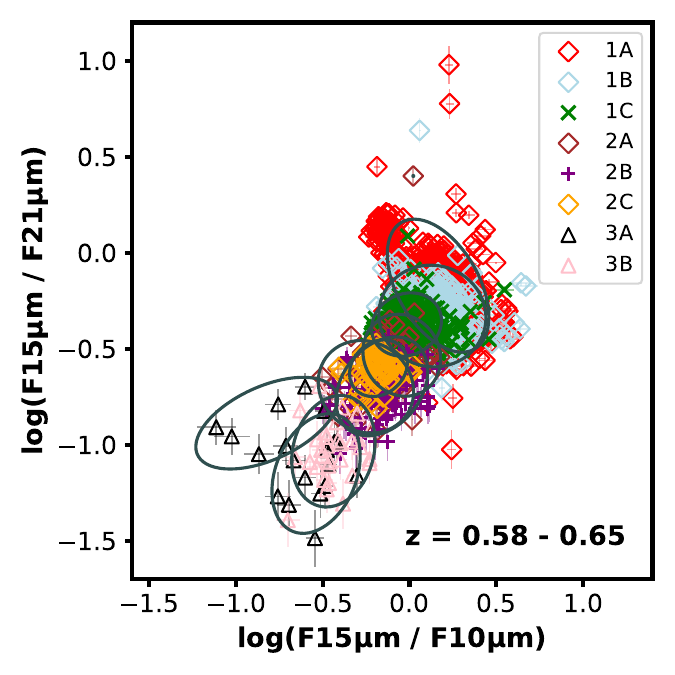}\\
\includegraphics[scale=0.5]{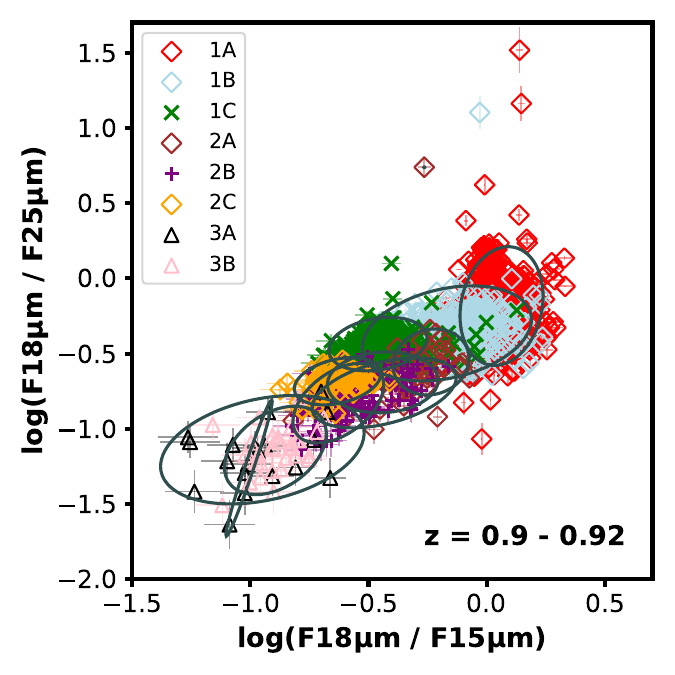}&
\includegraphics[scale=0.5]{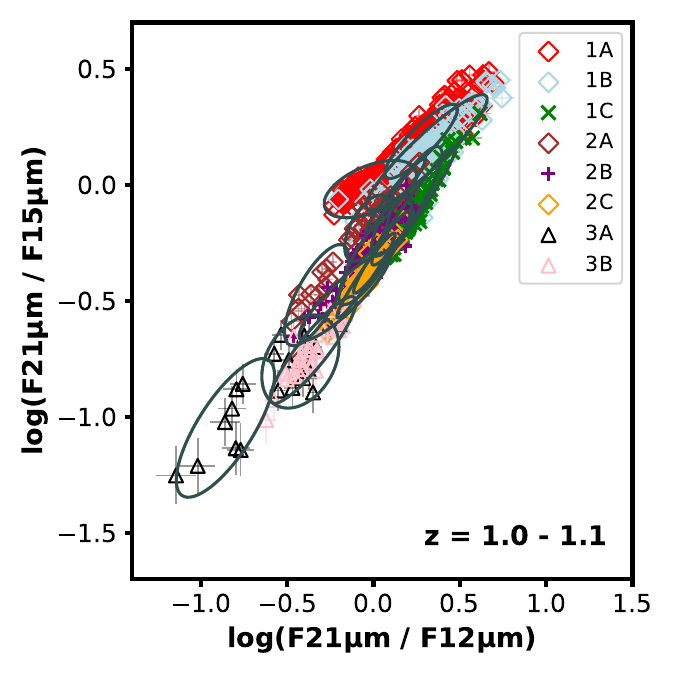}&
\includegraphics[scale=0.5]{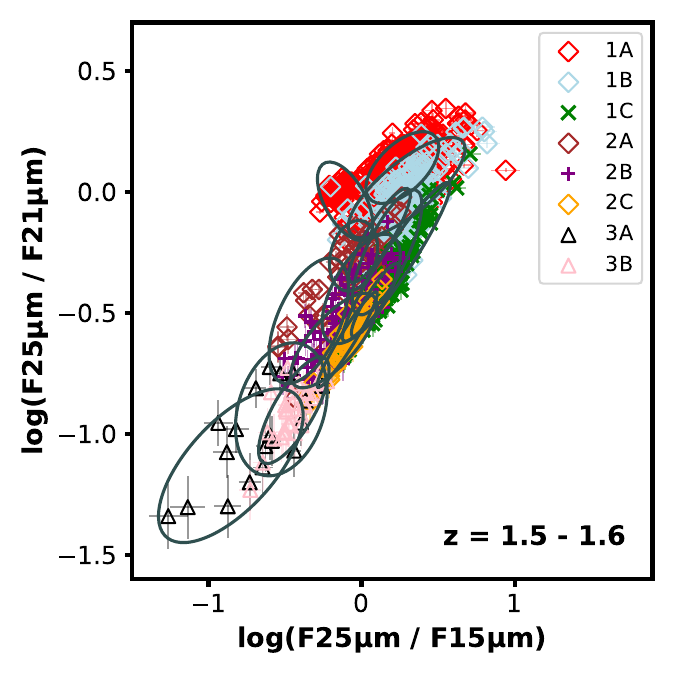}\\
\end{array}$
\end{center}
\caption{ MIRI colors of different MIR class  IDEOS galaxies at $z=$ 0.25--0.30 (top left panel), $z=$ 0.58--0.65 (top right and middle left panels), $z=$ 0.90--0.92 (middle right panel), $z=$ 1.00--1.10 (bottom left panel), $z=$ 1.50--1.60 (bottom right panel). The MIR class symbol code is given in the legend. In each panel, the ellipses represent a 2D Gaussian component for the MIR class subregions. 
}
\label{fig:fig1}
\end{figure*}

First, we used the Gaussian mixture modeling \citep[GMM,][e.g.,]{Maas15} to determine the regions where different MIR classes can be distinguished from each other. GMM is a clustering algorithm to model overlapping and not well-separated clusters by assuming that the data are created from a mixture of Gaussian distributions. We used the Python package \textsc{sklearn.mixture} from \textsc{scikit-learn} \citep[][]{Pedregosa11a}. We applied GMM to each MIR class by allowing one, two, or three separate Gaussian distributions, each with its mean, covariance, and weight parameters. Since our color--color diagrams are in a 2D space, the mean is a 2D vector and the covariance is a 2x2 matrix. The ellipses in the panels of Fig. \ref{fig:fig1} show the obtained model regions for each MIR class; model regions are tabulated in Table \ref{t1} in Appendix \ref{S:gmmfit}. 

In all panels in Fig. \ref{fig:fig1}, a distinct color region of class 3A can be seen in the lower left corner, which does not overlap with other class regions. This is caused by the large flux difference between Si97 and PAH62, and PAH77 that is expected to be large for absorption-dominated galaxies. For example, at $z=$ 0.25--0.30, PAH62 is at F770W, PAH77 is at $F1000W$, Si97 is at $F1280W$. Therefore, the $\rm{F}(12\,\mu$m)$/\rm{F}(7\,\mu$m) and $\rm{F}(12\,\mu$m)$/\rm{F}(10\,\mu$m) measure the flux difference between Si97 and PAH62, and PAH77, respectively. At $z=$  0.58--0.65 range, PAH62 is at $F1000W$, PAH77 is at $F1280W$, PAH113 is at $F1800W$, PAH127 is at $F2100W$, and Si97 is at $F1500W$. And at $z=$  0.9--0.92 range, PAH77 is at $F1500W$, PAH127 is at $F2550W$, Si97 is at $F1800W$. Similarly, at $z=$  1.0--1.1 range, PAH62 is at $F1280W$, PAH77 is at $F1500W$, Si97 is at $F2100W$. And at $z=$  1.5--1.6 range, PAH62 is at $F1500W$, PAH77 is at $F2100W$, Si97 is at $F2550W$.
As shown in each panel in Fig. \ref{fig:fig1}, for each $z$ range examined here, similar regions are detected for galaxies 3A and 3A-3B classes due to the large flux difference between Si97 and PAH lines. In most panels, 2A galaxies with relatively smaller flux difference between Si97 and PAH lines locate above the absorption-dominated galaxies.

AGN produce a strong continuum from hot dust, but no PAH emission. For example, at $z=$ 0.25--0.30 continuum regions 6--6.9\,$\mu$m, 7--8.3\,$\mu$m, and 9.5--11.1\,$\mu$m  are at F770W, $F1000W$, and $F1280W$, respectively. For (unabsorbed) AGN, the $\rm{F}(12\,\mu$m)$/\rm{F}(7\,\mu$m) and $\rm{F}(12\,\mu$m)$/\rm{F}(10\,\mu$m) represent the relatively small flux difference between the two continua.  Therefore, 1A and 1A-1B AGN are located at similar regions at the top in each panel due to the small flux difference in the colors on the y-axis. 

Here we find that at different $z$ ranges, the MIR class region boundaries are similar when obtained from the ratios of MIRI bands where Si and the same PAH lines shift. We note that our attempt to define distinct regions for each MIR class, is mostly successful for the 1A, 3A, and 3A-3B classes. However, for other classes the defined regions heavily overlap with each other.

\subsubsection{Classification of AGN, SFGs, and composite galaxies}\label{S:onlyagnsfg}
In addition to the defined distinct regions designated to select only a single MIR class in the GMM analysis, we combine 1A-1B-2A as AGN, 1C-2C as SFGs to separate AGN and SFGs. Modeled AGN (orange), SFGs (green) region ellipses are shown in Fig. \ref{fig:newfig2}. The modeled regions are listed in Table \ref{t2} in Appendix \ref{S:gmmfit}.

\begin{figure*}
\begin{center}$
\begin{array}{lll}
\includegraphics[scale=0.5]{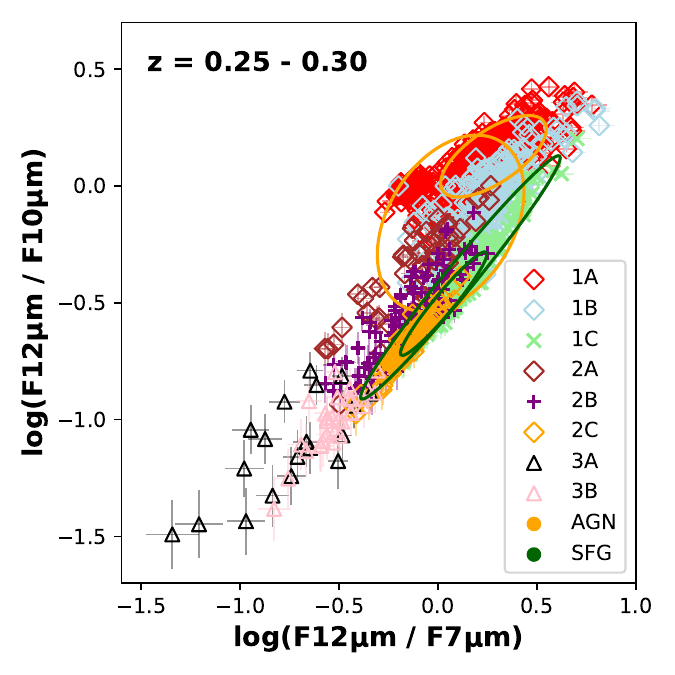}&
\includegraphics[scale=0.5]{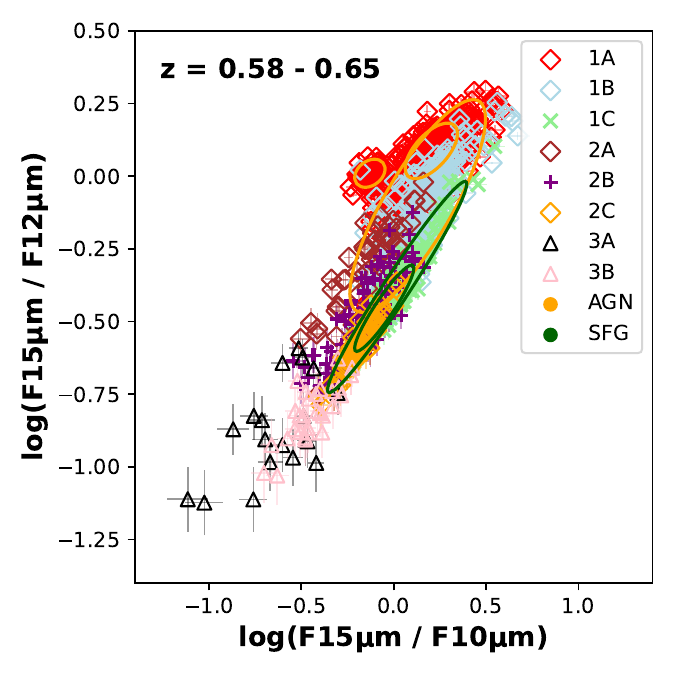}&
\includegraphics[scale=0.5]{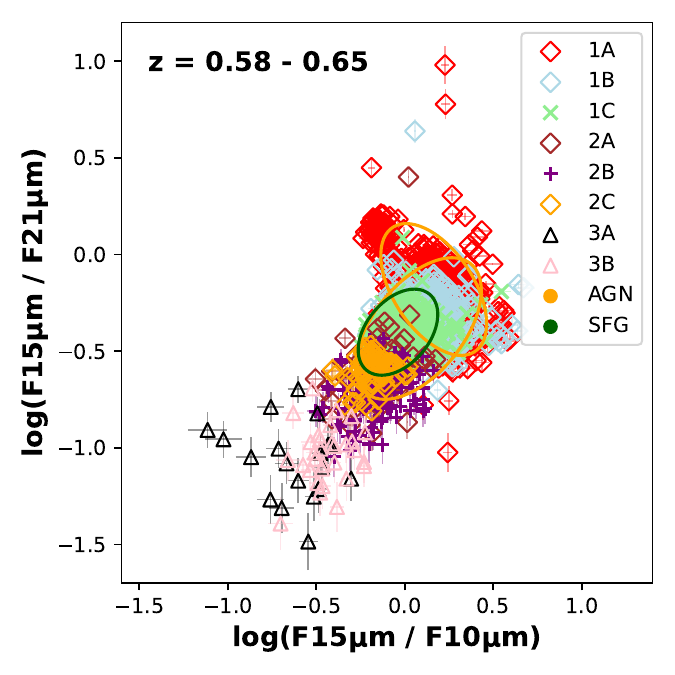}\\
\includegraphics[scale=0.5]{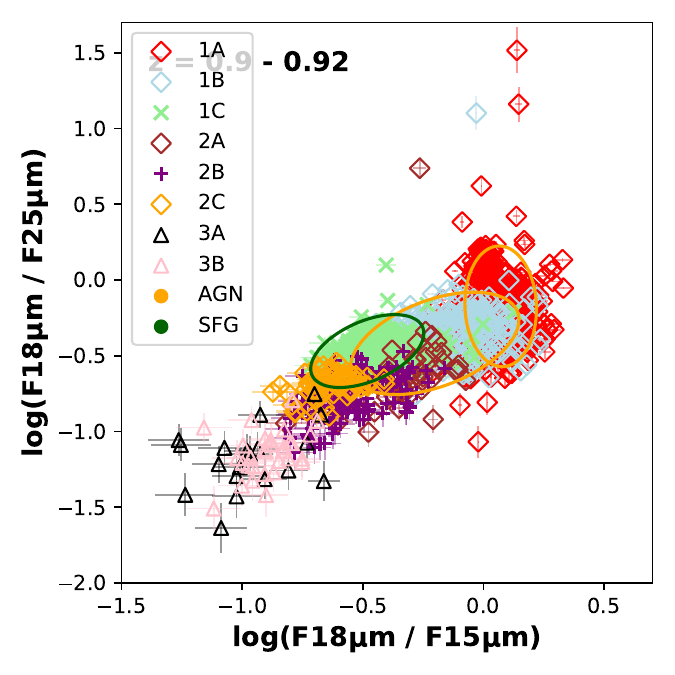}&
\includegraphics[scale=0.5]{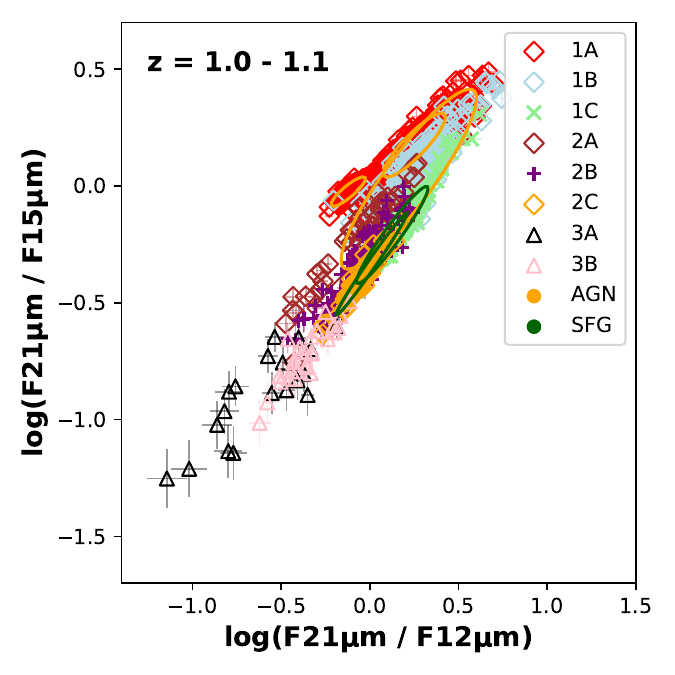}&
\includegraphics[scale=0.5]{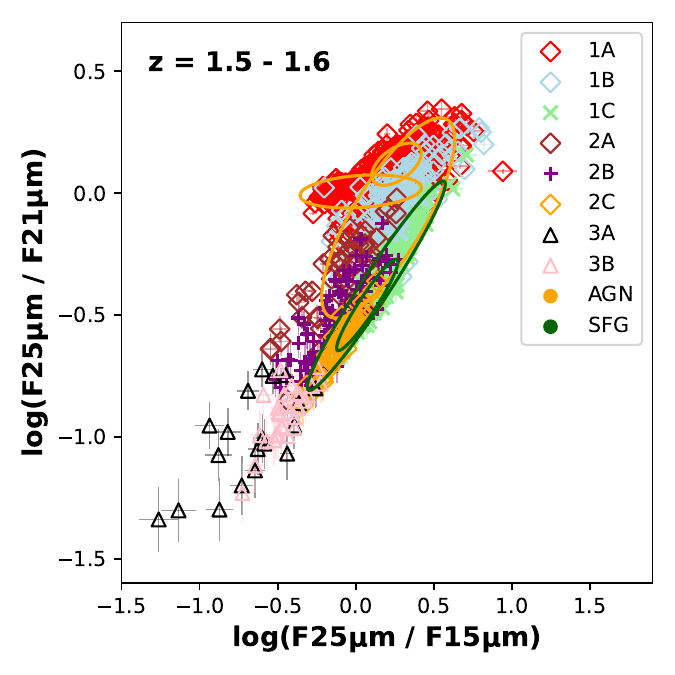}\\
\end{array}$
\end{center}
\caption{ Subregions obtained from the GMM analysis for AGN (1A-1B-2A combined) and  SFGs (1C-2C combined). The regions of AGN and SFGs are shown as orange and green ellipses, respectively. 
}
\label{fig:newfig2}
\end{figure*}

\subsection{Comparison with MIRI color classifications in the literature}\label{S:result3}

\begin{figure*}
\begin{center}$
\begin{array}{lll}
\includegraphics[scale=0.5]{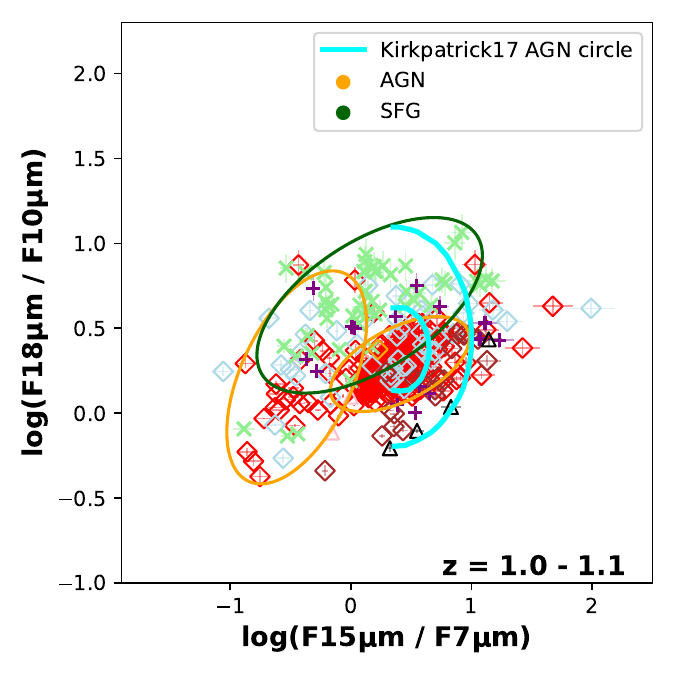}&
\includegraphics[scale=0.5]{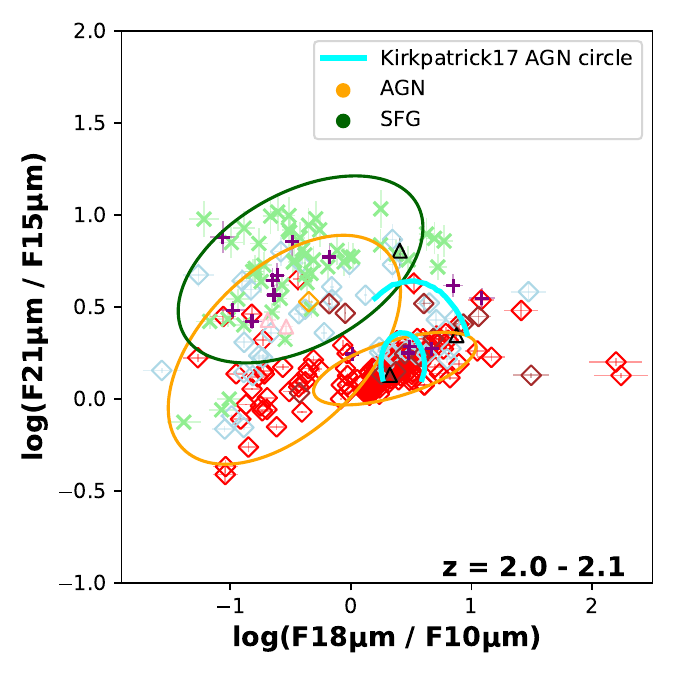}&
\includegraphics[scale=0.5]{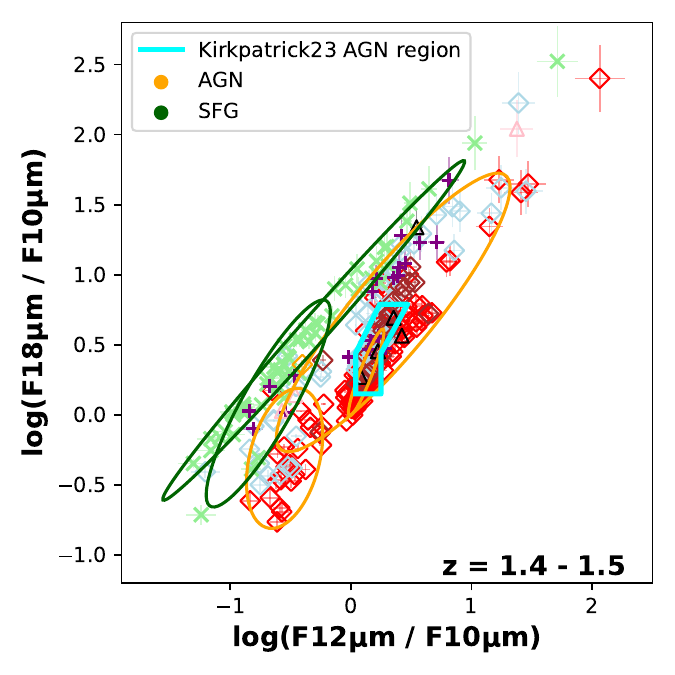}\\
\end{array}$
\end{center}
\caption{ Left and middle panels: Color--color diagrams adopted from \citet[][]{Kirkpatrick2017b}. The solid cyan circles are their AGN, composite, and SFGs regions. Right panel: Color--color diagram \citet[][]{Kirkpatrick2023}. The cyan hexagon defines their AGN region. In all panels, the dark green and orange ellipses respectively represent the SFGs and AGN regions obtained by the GMM analysis.} 
    \label{fig:newfig3}
\end{figure*}

\citet[][]{Kirkpatrick2017b} and \citet[][]{Kirkpatrick2023} present redshift-dependent MIRI color selection for AGN, SF and composite galaxies. 
Unlike our study, their galaxy classification is based on SED templates, and galaxies are classified according to their AGN MIR contribution in the $5-15\,\mu$m wavelength range ($f_{\rm AGN}^{5-15\mu m}$). They define AGN as $f_{\rm AGN}^{5-15\mu m}\geq 0.7$, SFGs as $f_{\rm AGN}^{5-15\mu m}<0.3$ and composites as $0.3\leq f_{\rm AGN}^{5-15\mu m}<0.7$.
In Fig. \ref{fig:newfig3} we show the color--color diagrams adopted from \citet[][]{Kirkpatrick2017b} (left and middle panels). The inner solid cyan circle shows the AGN circle defined by \citet[][]{Kirkpatrick2017b}. Most of the 1A AGN in our sample are located in this region, but there are also composite AGN+SFGs. Between the inner and outer circles and outside the outer circle, we have all types of galaxies, AGN, SF, and composites.  
As noted by \citet[][]{Kirkpatrick2023} the color selection in these two diagrams is not good enough to select AGN in MIRI surveys such as CEERS. Therefore, they introduced an updated color--color diagram, shown in the right panel of Fig. \ref{fig:newfig3}. Here they define the AGN region with the cyan hexagon. Our sample has 1A-1B-2A AGN, 3A, and 2B galaxies inside their hexagon. 
In these diagrams, AGN and SFGs regions obtained by GMM analysis are shown by orange and dark green ellipses.
These regions are listed in Table  \ref{t3} in Appendix \ref{S:gmmfit}.

The AGN and SFGs limits of \citet[][]{Kirkpatrick2017b} and \citet[][]{Kirkpatrick2023} are inside the AGN and SFGs regions obtained by the GMM analysis; therefore, our regions are compatible with their limits. Compared to these studies, we determined regions in these color diagrams where galaxies of different MIR spectral types can be more accurately distinguished.

\section{MIR spectral classification of MIR detected galaxies in  SMILES}\label{S:smiles}

\subsection{SMILES survey }\label{S:smilessurvey}

The Systematic Mid-infrared Instrument Legacy Extragalactic Survey \citep[SMILES;][]{Alberts2024,Rieke2024} is a MIRI imaging survey of 34 arcmin$^{2}$ in the Great Observatories Origins Deep Survey South (GOODS-S) Hubble Ultra Deep Field \citep[HUDF;][]{Beckwith2006} field. It includes the $F560W$, $F770W$, $F1000W$, $F1280W$, $F1500W$, $F1800W$, $F2100W$, $F2550W$ MIRI broad bands. SMILES is the largest, deepest MIRI survey with superb spatial resolution (between 0.19 to 0.86 arcsec) ever completed. Therefore, it provides an ideal data set for our study.

SMILES survey area also includes ancillary data, including ultradeep NIRCam imaging from the JWST Advanced Deep Extragalactic Survey \citep[JADES;][]{Eisenstein2023a,Rieke2023}, JWST Extragalactic Medium-band Survey \citep[JEMS;][]{Williams2023b,Rieke2023}, Hubble Space Telescope (HST) imaging from multiple surveys \citep[e.g.,][]{Grogin2011}, NIRCam grism spectroscopy from the First Reionization Epoch Spectroscopically Complete Observations \citep[FRESCO;][]{Oesch2023}.

\subsection{SMILES data }\label{S:smilesdata}

We used the SMILES Data Release 1 (DR1) clean catalog \citep[][]{Alberts2024} with 3096 sources detected at $\geq 4 \sigma$ in either the $F560W$ or $F770W$ bands. We refer to \citet[][]{Alberts2024} for the details of the photometric measurements.
The catalog provides aperture-corrected photometry in five circular apertures and in a 2.5$\times$-scaled Kron aperture. In this work, we used 2.5$\times$-scaled Kron aperture measurements as recommended by \citet{Rieke2023}.

To apply our $z$-dependent color selection obtained in this work we required $z$ determinations from the ancillary data and the literature.
We matched the SMILES DR1 catalog with JADES Data Release 3 (DR3) catalog\footnote[1]{jades$\_$dr3$\_$medium$\_$gratings$\_$public$\_$gs$\_$v1.1.fits} \citep[][]{DEugenio2024} 
within 0.3 arcsec separation, to obtain spectroscopic redshifts ($z_{s}$) of 286 SMILES sources (from their $z\_Spec$ column). 
We matched SMILES DR1 catalog with MUSE Hubble Ultra Deep Field surveys DR 2 Main source catalog \citep[][]{Bacon2023} within 0.6 arcsec separation, to obtain $z_{s}$ of 594 SMILES sources (from their $z$ column). 
We matched SMILES DR1 catalog with CANDELS GOODS-S Redshift catalog \citep[][]{Kodra2023} 
within 0.6 arcsec to obtain $z_{s}$ of 467 SMILES sources. In total, we obtained spectroscopic redshifts for unique 1101 sources (from their $z\_spectroscopic$ column). 
We matched the SMILES DR1 catalog with JADES NIRCam photometric redshift catalog\footnote[2]{hlsp$\_$jades$\_$jwst$\_$nircam$\_$goods-s-deep$\_$photometry$\_$v2.0$\_$catalog.fits}
\citep[][]{Rieke2023} within 0.3 arcsec and obtained the photometric redshifts ($z_{p}$) of 2864 sources obtained by the template-fitting code EAZY \citep[][]{Brammer2008} to the JADES and JEMS JWST/NIRCam and HST/ACS photometry (their $EAZY\_z\_a$ column). 
 Additionally, since we used SMILES AGN sample of \citet{Lyu2024} in \S \ref{S:agns} for 201 overlapping SMILES sources (within 0.6 arcsec positional match), we used their revised photometric/spectroscopic redshifts compared to CANDELS, JADES and JEMS catalogs.

\section{MIRI color classification of SMILES sample}\label{S:colsmiles}

We performed classification of SMILES sources by taking into account both the colors and the $z$ of the sources (for $z_{p}$ values we required $z_{p}-1\sigma$ and $z_{p}+1\sigma$ to be within the applied $z$ range). We used GMM models (listed in Tables \ref{t1}, \ref{t2}, and \ref{t3}) trained on IDEOS MIR classes. 

We used both Gaussian Mixture Model prediction and Mahalanobis distance minimization for initial classification. Gaussian Mixture Model prediction predicts the class of a source based on maximum likelihood assignment. Alternatively, Mahalanobis distance criterion assigns each object to the closest GMM component. While both approaches assign sources to the same dominant classes in most cases, discrepancies arise for outlier points close to decision boundaries. Mahalanobis distance emphasizes proximity to class centroids while incorporating shape and orientation of the class ellipses. GMM prediction, in contrast, may assign outliers to components with broader covariance structures. While both methods yield largely consistent classifications, the Mahalanobis approach tends to provide more conservative assignments, especially for sources lying at the intersection of multiple ellipses. Therefore, we calculated Mahalanobis distance between each data point and each GMM component for classification of SMILES sources. We used 95.4\% confidence as a threshold to assign the MIR class based on highest confidence. 

\begin{figure*}
\begin{center}$
\begin{array}{lll}
\includegraphics[scale=0.5]{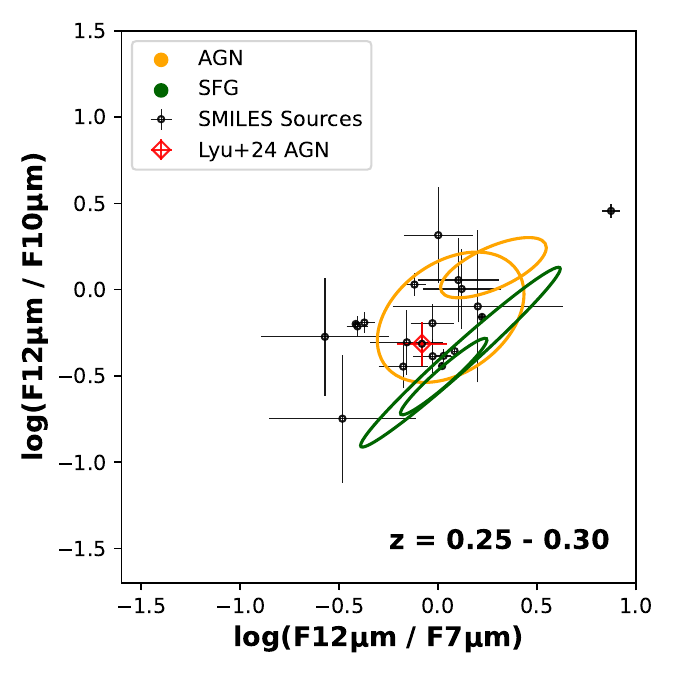}&
\includegraphics[scale=0.5]{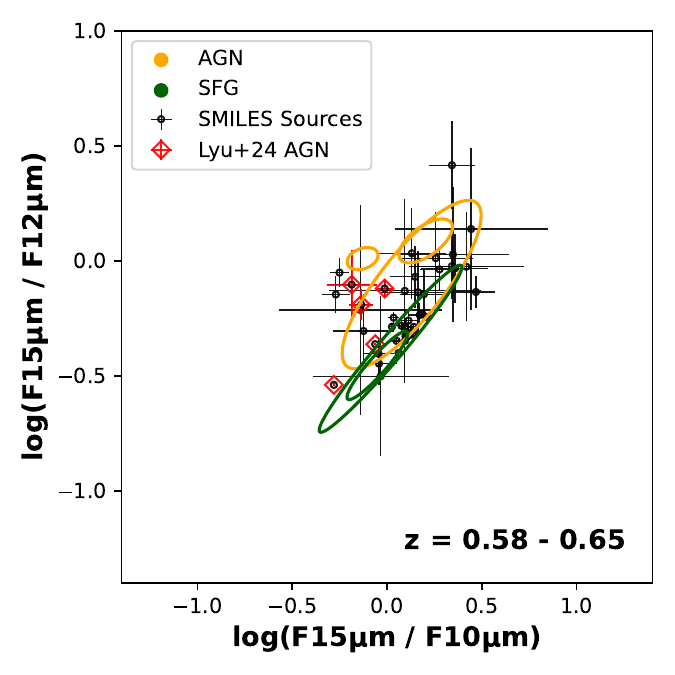}&
\includegraphics[scale=0.5]{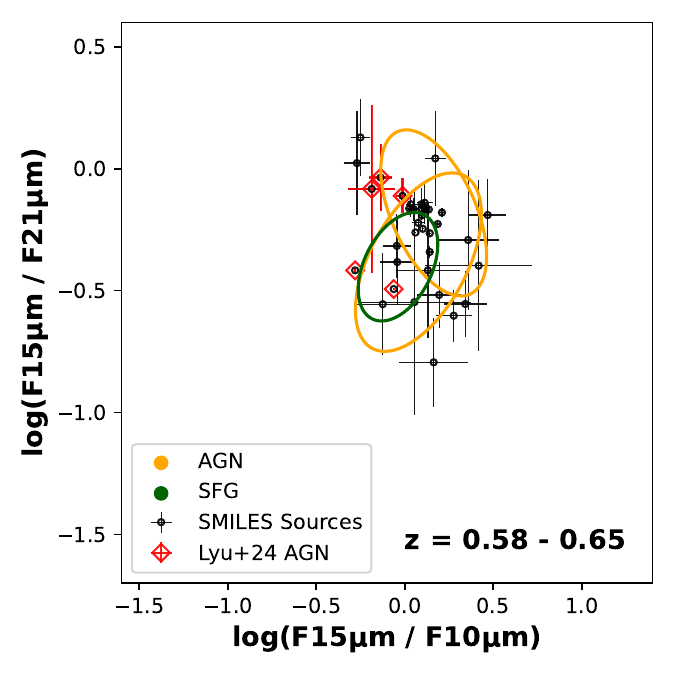}\\
\includegraphics[scale=0.5]{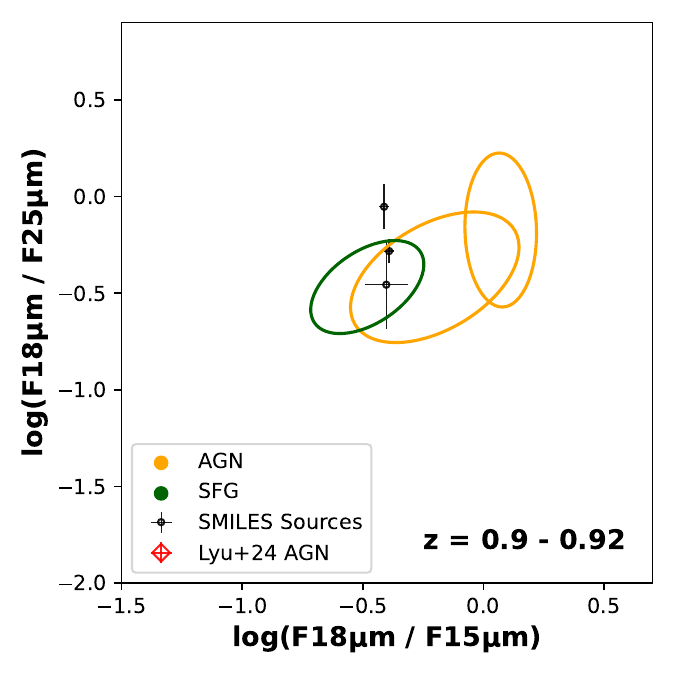}&
\includegraphics[scale=0.5]{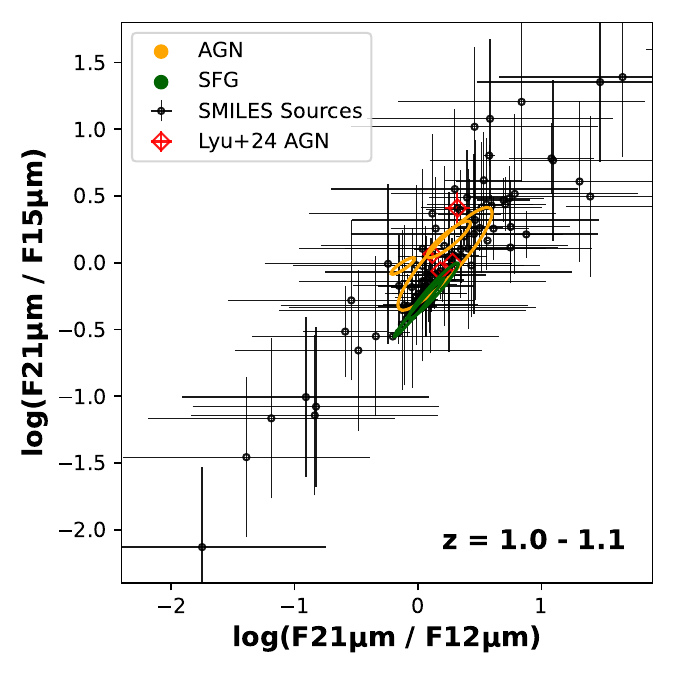}&
\includegraphics[scale=0.5]{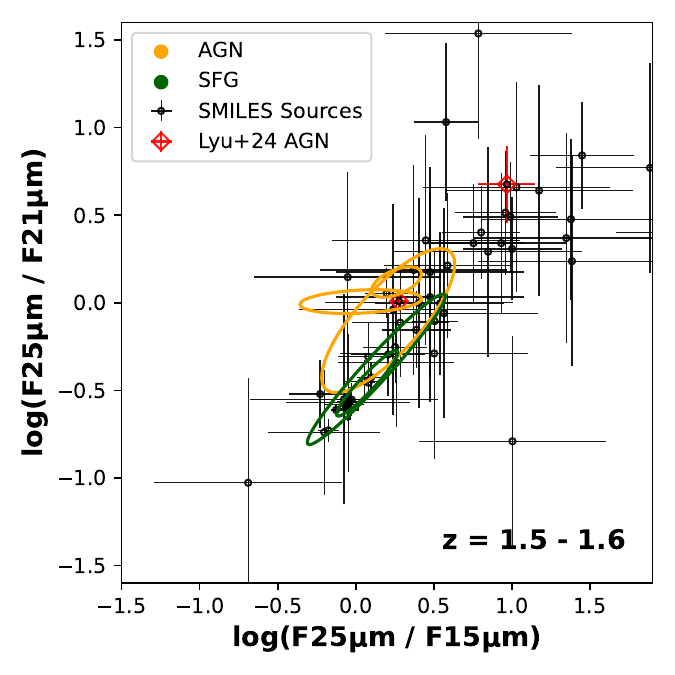}\\
\includegraphics[scale=0.5]{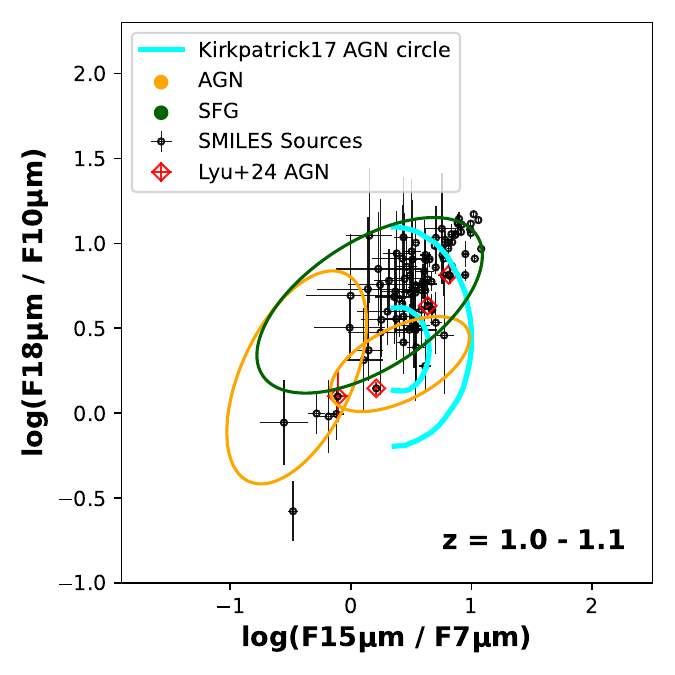}&
\includegraphics[scale=0.5]{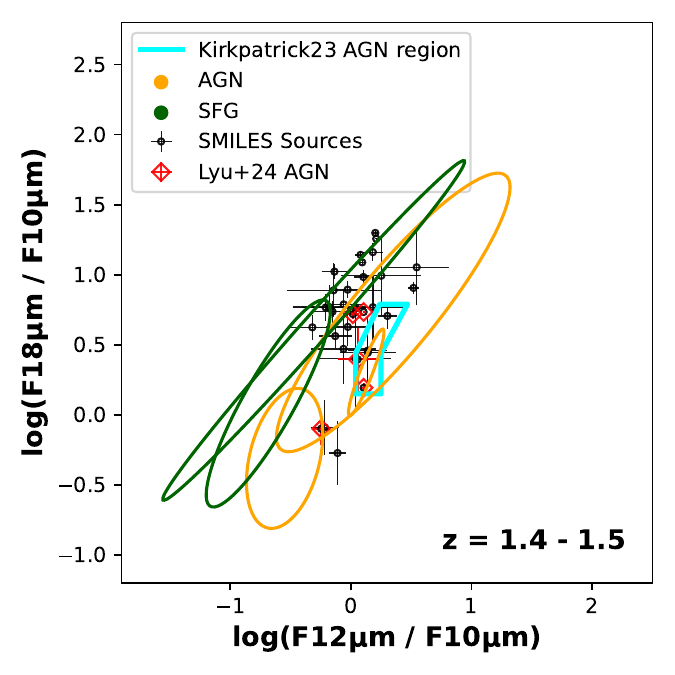}&
\includegraphics[scale=0.5]{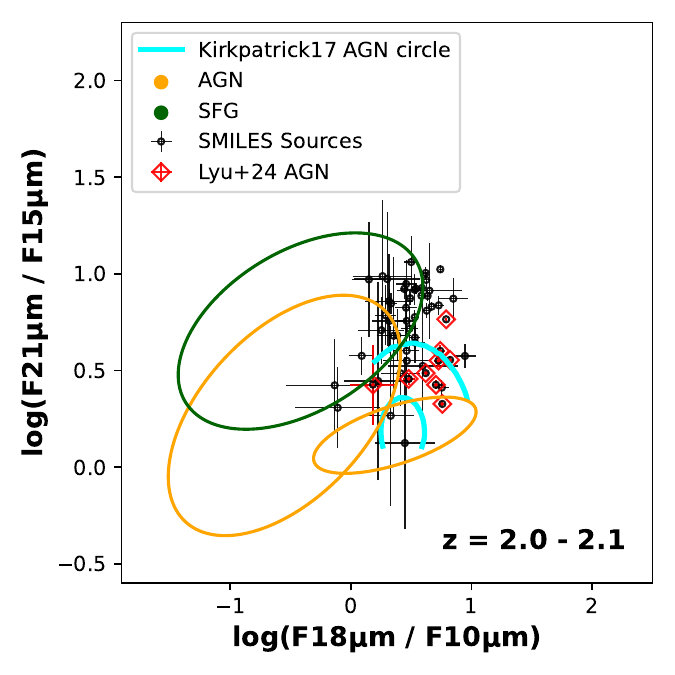}\\
\end{array}$
\end{center}
\caption{ MIRI colors of SMILES galaxies.
The subregions obtained from the GMM analysis for AGN (1A-1B-2A combined), SFGs (1C-2C combined) are shown as the orange and dark green ellipses.  
}
    \label{fig:fig3}
\end{figure*}

All of the $z-$dependent applied color limits are shown in Fig. \ref{fig:fig3}.   
We identified 121  AGN (including 1A, 1B, 2A), 154  SFGs (including 1C, 2C), and six Si absorption-dominated galaxies (3A class), at various redshifts up to $z\ge2$. 

% NEW Fig. 4

\section{SED analysis of the selected SFGs}\label{S:seds}

To investigate the nature of the 154 MIRI-selected SFGs we combined the available photometric data and performed SED analysis with the Code for Investigating Galaxy Emission \citep[CIGALE;][]{Noll2009,Serra2011,Boquien2019,Yang2022} version 2022.1 \citep{Yang2022}. 
We used 31 filters in the SED fitting: nine HST filters (the HST/Advanced Camera for Surveys (ACS) $F435W$, $F606W$, $F775W$, $F814W$, $F850LP$ and HST/WFC3 $F105W$, $F125W$, $F140W$, $F160W$) from \citet{Rieke2023}; 14 NIRCam wide and medium filters ($F090W$, $F115W$, $F150W$, $F182M$, $F200W$, $F210M$, $F277W$, $F335M$, $F356W$, $F410M$, $F430M$, $F444W$, $F460M$, and $F480M$) from  \citet{Rieke2023}; eight MIRI filters ($F560W$, $F770W$, $F1000W$, $F1280W$, $F1500W$, $F1800W$, $F2100W$, and $F2550W$) from \citet{Alberts2024}. We used Kron magnitudes as recommended by \citet{Rieke2023}. We checked the flags indicating a star (FLAG$\_$ST), a nearby bright star (FLAG$\_$BS) and contaminating the photometry (FLAG$\_$BN) and used only the reliable photometry from the catalogs of \citet{Rieke2023}. Additionally, when available, we used  \textit{Spitzer} MIPS photometry at 70\,$\mu$m and \textit{Herschel} PACS photometry at 100 and 160\,$\mu$m for the 0.6 arcsec matched counterparts from the GOODS-\textit{Herschel} catalog \citep{Elbaz2011}. However, only 29 SFGs and 15 AGN have reliable  \textit{Herschel} PACS photometry.

In addition, the FIR radiation, which is a strong indicator of SF in galaxies, has been detected \citep[in the GOODS-\textit{Herschel}catalog][]{Elbaz2011} for 29 sources. Hence, at least 19 percent of the SFG candidates are secure SFGs.

\begin{figure*}
\begin{center}$
\begin{array}{lll}
\includegraphics[scale=0.6]{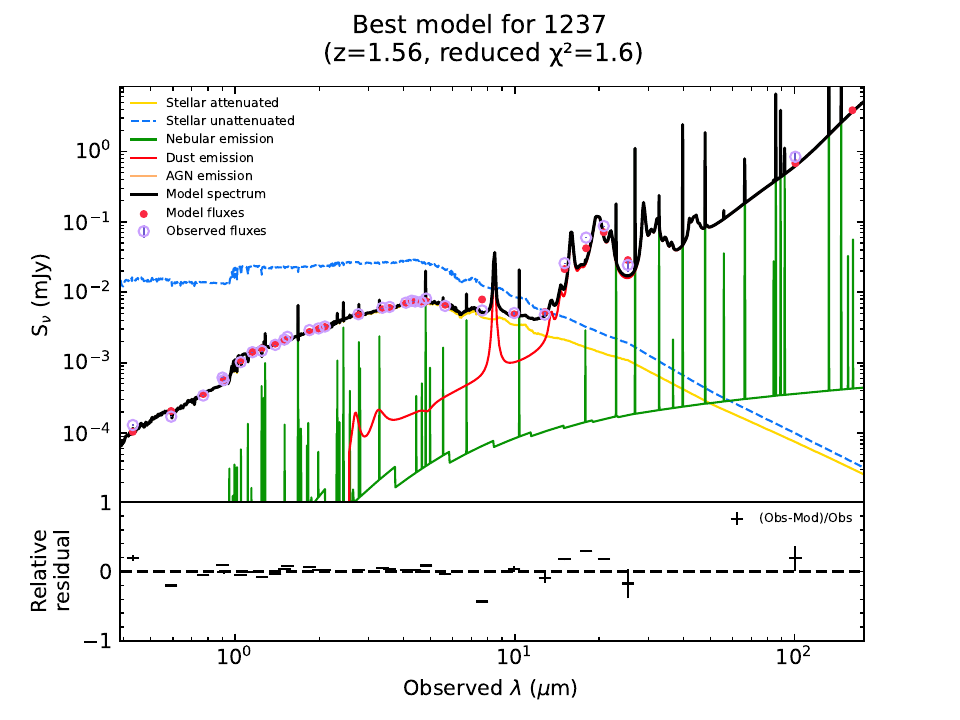}&
\includegraphics[scale=0.6]{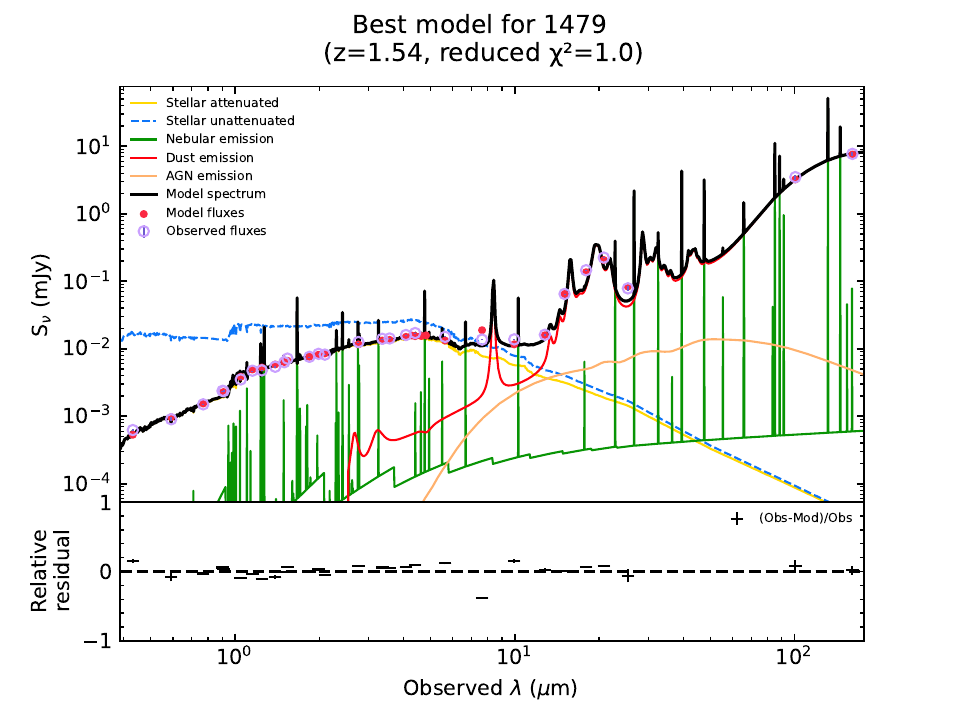}\\
\end{array}$
\end{center}
\caption{ Two examples (SMILES IDs 1237 and 1479) of the best-fitting SEDs of the MIRI-selected SFGs candidates at $z\sim1.5$. The red and blue curves  show the dust emission and the stellar emission, respectively. The purple circles show the observed photometric fluxes. The filled red circles are the model fluxes. The best model for 1479 requires an AGN component; in 1237 there is no AGN.}
    \label{fig:newfig5}
\end{figure*}

CIGALE\footnote[3]{\url{http://cigale.lam.fr/}} is a SED analysis code designed especially for galaxies producing IR radiation (from near to far-IR). 
It combines the radiation produced by the stars, star formation, dust, and AGN by balancing \citep[e.g.,][]{ Efstathiou2003, daCunha2008, Noll2009, Boquien2019} the ultraviolet/optical/near-IR radiation and the re-emitted mid- and far-IR radiation by dust. 
We used the CIGALE modules listed in Table \ref{t4} for different emission components contributing to the SED. 
Since all of the SFG candidates have $z_{s}$ we fixed the source redshift in the SED analysis.
To model the stellar component we used the delayed SFH optional exponential burst module \citep{Boquien2019}.
We used the stellar population models synthesis models of \citet{Bruzual2003} and the Salpeter initial mass function (IMF) \citep{Salpeter1955} with solar metallicity. We adopted the dustatt$\_$modified$\_$starburst \citep{Leitherer2002} that assumes the extinction law of \citet{Calzetti2000} for the dust attenuation. We allowed the color excess of nebular lines $E(B-V)_{line}$ to vary between 0.05 and 4. We fixed the reduction factor for the stellar continuum attenuation as 0.44.
We used the \citet{Draine2014} updated models for the dust emission and allowed a large range for the minimum radiation field (0.4 $\le U_{min} \le$ 50). We allowed the mass fraction of PAH to be between 1.77 and 7.32. We used the clumpy torus AGN model of \citep{Stalevski2012, Stalevski2016} for the AGN component.  We allowed all possible values of the average edge-on optical depth at 9.7 $\mu$m. We also allowed to have Type 1 and Type 2 AGN based on all available viewing angles. The fractional AGN emission contribution to the IR emission in the 8$-$1000 $\mu$m range is shown by the $frac_\mathrm{AGN}$ parameter. To account for possible AGN strengths we allowed $frac_\mathrm{AGN}$ to be between 0 and 0.7.   
We show two examples of the best-fitting SEDs of the MIRI-selected SFGs at $z\sim1.5$ in Fig. \ref{fig:newfig5}. 
 
CIGALE uses the given parameters and builds several model combinations \citep[e.g.,][]{Boquien2019}. The Bayesian method used in CIGALE measures the likelihood of each model compared to the dataset. CIGALE uses these likelihoods to give a probability distribution function of the parameters and provides the Bayesian uncertainties based on Bayesian analysis. 

\begin{table*}
\caption{\label{t4} \texttt{CIGALE} modules}
\centering
\begin{tabular}{cc}
\hline
Parameters & Value \\
\hline
             \multicolumn{2}{c}{\textit{Delayed SFH with optional exponential burst} \citep{Boquien2019}} \\
             \hline
             e-folding time of the main stellar population [$10^6$ yr] & 500, 1000.0, 4000.0, 5000.0 \\
             Age of the galaxy's main stellar population [$10^6$ yr] & 500, 1000.0, 2000.0, 5000.0 \\
             e-folding time of the late starburst population [$10^6$ yr] & 50.0 \\
             Age of the late burst [$10^6$ yr] & 10.0 \\
             Mass fraction of the late burst population & 0.00 \\
             multiplicative factor controlling the amplitude of SFR if normalization is True & 1.0 \\
             \hline
             \multicolumn{2}{c}{\textit{SSP} \citep{Bruzual2003}} \\
             \hline
             Initial mass function & \cite{Salpeter1955} \\
             Metallicity & 0.02 \\
             Age of separation between the young and old star populations & 10.0 \\
             \hline
             \multicolumn{2}{c}{\textit{Dust attenuation} \citep{Charlot2000}} \\
             \hline
             Logarithm of the V-band attenuation in the ISM &  0.1, 0.17, 0.28, 0.46, 0.77, 1.29, 2.15, 3.59, 5.99, 10.0\\
             Ratio of V-band attenuation from old and young stars & 0.44 \\
             Power-law slope of the attenuation in the ISM & -0.7,-0.5 \\
             Power-law slope of the attenuation in the birth cloud & -0.7\\
             \hline
             \multicolumn{2}{c}{Dust emission \citep{Draine2014}} \\
             \hline
             Mass fraction of PAH (q$_{PAH}$) & 1.12, 2.5,  3.9, 4.58, 5.95, 6.63, 7.32 \\
             Minimum radiation field (${\rm U}_{\rm min}$) & 0.1, 1.0, 10.0, 20.0, 50.0 \\
             Power-law slope $\alpha$ ($\frac{{\rm dU}}{{\rm dM}} \propto U^\alpha$) & 2.0, 3.0 \\
             Fraction illuminated from ${\rm U}_{\rm min}$ to ${\rm U}_{\rm max}$ & 0.01, 0.1 \\
             \hline
             \multicolumn{2}{c}{SKIRTOR AGN} \\
             \hline
             $\tau_\mathrm{AGN}$    & 3\\
             $oa_{\rm{AGN}}$    & 40, 70 \\
		$R_{\rm{AGN}}$    &  20 \\
		inclination ($i$)                          & 0, 10,  30,  80 \\
		$frac_\mathrm{AGN}$ (computed for the 3-30 micron range)            &  0.0, 0.01, 0.03, 0.05, 0.1, 0.2, 0.3, 0.4, 0.5, 0.6\\
             \hline
        \end{tabular}
\end{table*}

\section{Results and discussion}\label{S:results} 

\subsection{MIRI color selection}\label{S:col} 

The most important limitation of the color criteria proposed in this study is that they cannot select all galaxies belonging to a specific MIR spectral class. 
On the other hand, our color criteria have the advantage of selecting sources that can be distinguished from other classes. This purity is the most important difference from the color limit studies in the literature. Since our color limits are valid in small $z$ intervals, we expect to have a low level of confusion between SFGs and AGN compared to previous studies \citep[e.g.,][]{Kirkpatrick2023}.

Since our color limits are based on the shift of the spectral lines through the MIRI bands, the proposed limits are redshift-dependent. Therefore, prior $z$ knowledge is a second limitation of their application. Therefore, we suggest using these limits with secure $z_{s}$ or $z_{p}$ with low uncertainties. 

Possible biases in selecting AGN, SFGs, and Si absorption-dominated galaxies by MIR colors include underrepresentation of each galaxy class, as they overlap in color space with other MIR classes. MIR color selection is sensitive to AGN-dominated emission, leading to biases, particularly mergers where extreme nuclear starbursts can mimic AGN signatures. While MIR color selection primarily identifies luminous, obscured AGN, it potentially overlooks starbursts that may also exhibit a similar signature. These selection biases can lead to misclassification that can affect the completeness of AGN and SFGs identified by these diagnostics.

\subsection{MIRI selected AGN toward cosmic noon}\label{S:agns}

We performed SED analysis for the 121 MIRI-selected AGN candidates using the modules listed in Table \ref{t4} with a more comprehensive parameter setting for the AGN component (            $\tau_\mathrm{AGN}=$ 3, 5, 7, 9, 11; $oa_{\rm{AGN}}=$10, 20, 30, 40, 50, 60, 70, 80; $i=$0, 10, 20, 30, 40, 50, 60, 70, 80, 90; $frac_\mathrm{AGN}=$ 0.0, 0.01, 0.05, 0.1, 0.2,0.3,0.4,0.5,0.6,0.7,0.8,0.9). 
We find that 31 of the 121 AGN candidates can be classified as AGN with $frac_\mathrm{AGN} \ge 0.2$ based on their SEDs. As a result of the SED analysis, we consider  31 sources\footnote{(SMILES IDs: 34, 38, 74, 90, 144, 153, 186, 327, 369, 373, 375, 436, 502, 529, 714, 796, 886, 1036, 1081, 1130, 1300, 1401, 1447, 1506, 1661, 1671, 1721, 1798, 1926, 2267, 2384)}  as secure AGN selected by MIRI colors. For 36\% AGN candidates, the SED selection agrees with the MIRI-color selection. 
Among the 79 candidates with $frac_\mathrm{AGN} < 0.2$, 39 are dwarf galaxies with 
$M_{\rm{star}} < 10^{9.5}M_{\odot}$. As shown in a recent MIRI SED analysis-based AGN classification study by \citet{Lyu2024}, dwarf galaxies may appear to have MIR colors of AGN due to their reduced PAH and warmer dust emission. Therefore, 39 dwarf galaxies among 121 AGN candidates indicate a 32\% contamination rate, potentially causing misclassifications.  

To quantify the success rate of our AGN selection based on a different validation metric, we compare with the MIRI-selected AGN sample of \citet{Lyu2024} in the SMILES survey. Their AGN selection includes various selection methods, including SED analysis and multi-wavelength properties such as X-ray detection, radio counterpart, and variability. They have 26 AGN that satisfy the $z$ conditions we apply. The legend of Fig. \ref{fig:fig3} reflects the red triangles of \citet{Lyu2024} AGN; 
18 of these are overlapping with our AGN candidates. 
Our MIRI AGN limits have chosen 18 of their AGN. 
Based on this comparison, we estimate the success rate of our AGN selection to be at least 69 percent. Among the remaining eight AGN in \citet{Lyu2024} sample, seven of them lie in the composite region of the bottom left and right panels in Fig. \ref{fig:fig3} \citep{Kirkpatrick2017b}, and one lies in our SFGs region (top right panel in Fig. \ref{fig:fig3}). However, for this case, the SFG region overlaps with the AGN region, and the same source is classified as AGN in the middle panel; so this discrepancy is related to the overlapping boundary region for these colors.
\citet{Kirkpatrick2017b} diagrams (bottom left and right panels in Fig. \ref{fig:fig3}) are effective at isolating AGN via their red dust continuum. \citet{Lyu2024} AGN sample has been identified through the MIRI SED analysis, and some sources may appear near or within composite/SFGs region in \citet{Kirkpatrick2017b} diagrams, since low-luminosity, obscured, or geometrically diluted AGN broadband photometry is dominated by host galaxy star formation. For lower luminosity AGN that can not produce a red, power-law-like MIR slope the SED analysis would still isolate the subdominant AGN component even when the integrated colors cannot separate the AGN.
We also note that none of our Si absorption-dominated candidates are in their AGN sample. 

As reported by \citet{Kirkpatrick2023}, AGN regions in color diagrams can be contaminated by MIR weak galaxies, and this might also be possible for some of the candidates shown in Fig. \ref{fig:fig3}. Since MIR weak galaxies do not indicate a specific MIR class, it is not possible to assign them to a class in this work. Some of the AGN that are not in the sample of \citet{Lyu2024} might be such MIR weak galaxies.

\subsection{MIRI selected SFGs toward cosmic noon}\label{S:sfgs}

We obtained the total IR luminosity ($L_{IR}$), SFR, and stellar mass ($M_{star}$) of 154 SFGs candidates from the SED analysis. The AGN contribution to $L_{IR}$ ($frac_\mathrm{AGN}$, in the 3-30 micron range) is zero for 84 sources, $frac_\mathrm{AGN} \leq 0.05 $ for 74 sources and $frac_\mathrm{AGN}=0.1$ for 13 sources. The low AGN contribution values ($frac_\mathrm{AGN}\leq 0.1$)  confirm that these sources are SFGs. In addition, FIR radiation, which is a strong indicator of SF in galaxies, has been detected \citep[in the GOODS-\textit{Herschel}catalog][]{Elbaz2011} for 29 sources. Hence, at least 19 percent of the SFGs candidates are secure SFGs. 

Since AGN contribution is not zero for some sources, we compute $L_{IR}$ as the sum of the dust luminosity and AGN luminosity. Most of the SFGs galaxies in our sample are normal SFGs with $10^{8.0} L_{\sun}  < L_{IR}< 10^{12.4} L_{\sun}$. Among our SFGs candidates, there are 25 LIRGs\footnote{SMILES ID:8, 55, 73, 339, 481, 486, 623, 1032, 1071, 1146, 1179, 1182, 1205, 1237, 1253, 1279, 1301, 1304, 1479, 1484, 1501, 1681, 1773, 1788, 1809} and three ULIRGs\footnote{SMILES ID:443,1373,1451} at $ 0.9 \le z\le 1.6$ (4 of which are starbursts with SFR $>100$). And there are 20 LIRGs\footnote{SMILES ID:135, 222, 223, 261, 336, 343, 377, 378, 581, 709, 829, 864, 878, 1014, 1196, 1256, 1540, 1686, 1699, 1813} and three ULIRGs\footnote{SMILES ID:358,720,988} at $z\sim 2$ (6 of which are starbursts with SFR $>100$).

\citet{Lin2024} applied $\log(\rm{F}(15\,\mu$m)$/\rm{F}(10\,\mu$m$)>0.8$ color selection determined from SED templates to identify  PAH luminous SFGs at $z \sim 1$ in the  Cosmic Evolution Early Release Science (CEERS) MIRI survey. Our $\log(\rm{F}(15\,\mu$m)$/\rm{F}(10\,\mu$m$)>1.0$ criteria agree well with theirs. Their sample includes ten SFGs with total IR luminosities of $10^{10}$ $\sim$ $10^{11.5}$ $L_{\odot}$ at  $z \sim 1$. They report their sample of normal SFGs to be on the star formation main sequence (MS, a tight correlation between SFR and stellar mass) at $z \sim 1$. 

\begin{figure*}
\sidecaption
  \includegraphics[width=12cm]{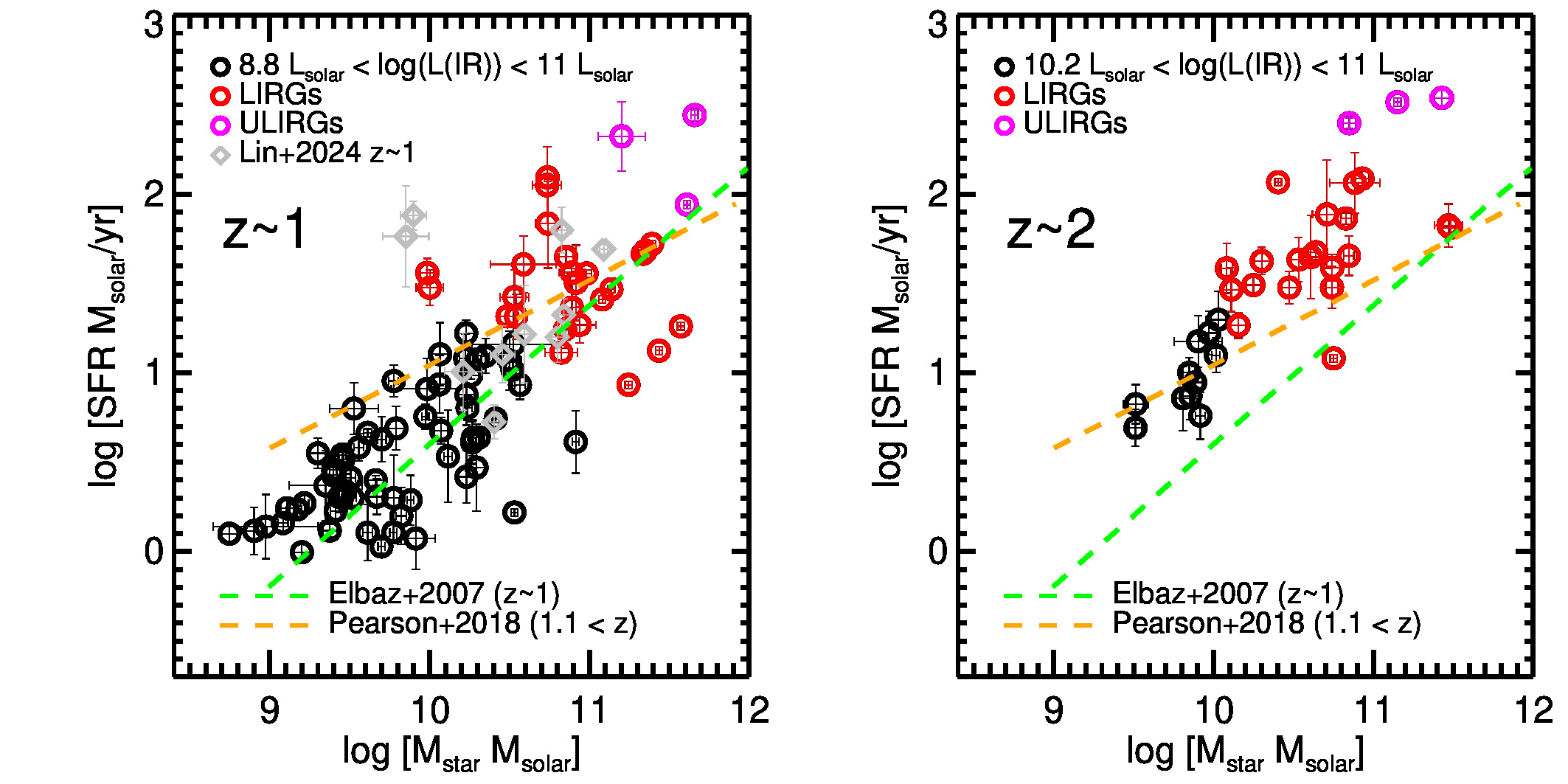}
     \caption{Left panel: Distribution of our $z\sim1$ MIRI-selected SFGs over the SFR--M$_{\odot}$ correlation at  $z\sim1$. The SFR and $M_{star}$ are obtained from the SED analysis. The green dashed line is $z=0.8-1.2$ MS relation of \citet{Elbaz2007}. The orange dashed line is $z=1.1-1.4$ the MS relation from \citet{Pearson2018}. The gray diamonds are the MIRI-selected SFGs sample of \citet{Lin2024}. The majority of our $z\sim1$ sample is consistent with the MS relation of \citet{Elbaz2007}. Right panel: Distribution of our MIRI-selected SFGs at $z\sim2$ over the SFR--M$_{\odot}$ correlation at $z\sim1$. The distribution of the $z\sim2$ SFGs indicates an evolution compared to $z\sim1$ MS.}
     \label{fig:newfig6}
\end{figure*}

The location of our SFGs sample of 90 galaxies ($ 0.9 \le z\le 1.6$) at $z \sim 1$ MS is shown in (the left panel of) Fig. \ref{fig:newfig6}. For comparison, we show the $z \sim 1$ SFGs sample of \citet{Lin2024} with gray diamonds. Our sample's $M_{star}$ spans a wider range than theirs. The MS relations for $z=1.1-1.4$ \citep{Pearson2018} and $z=0.8-1.1$ \citep{Elbaz2007} are shown by the dashed lines in Fig. \ref{fig:newfig6}.  Most of our $z \sim 1$ sample is distributed around the MS relation of \citet{Elbaz2007}. The locations of four (two LIRGs and two ULIRGs) $z \sim 1$ galaxies\footnote{SMILES ID:443,1182,1451,1479} above the MS relationships are compatible with starburst galaxies \citep[e.g.,][]{Daddi2010,Genzel2010}. In the left panel of Fig. \ref{fig:newfig6} 34 $z \sim 2$ SFGs are shown. Since their positions are above the dashed lines, their distribution indicates an evolution compared to the $z \sim 1$ MS.
As reported by \citet{Lin2024}, compared to previous IR space telescopes, JWST reveals less massive, lower IR luminosity SFGs at high-z; our MIRI selected SFGs sample supports this result with a larger number of galaxies.  

\subsection{MIRI selected silicate absorption-dominated galaxies toward cosmic noon}\label{S:3a3b}

The present MIRI color selection in this work is a promising tool to identify Si absorption-dominated galaxies at higher redshifts, up to $z \sim 1.6$. 
One of the most important results we have is the selection of Si absorption-dominated galaxies (3A) at high redshift. Our knowledge of \textit{Spitzer} detected 3A galaxies is limited to 32 galaxies at $z < 0.67$ and three galaxies at $z = 1.7 - 1.85$. Varying levels of silicate absorption, including the deep levels occasionally seen in IR-luminous low-$z$ galaxies, have
also been measured at higher redshifts ($z \sim 1 - 2$) \citep[e.g.,][and references therein]{Houck2005,Hatziminaoglou2015}. 
The selected 3A class candidates in this work might be the first JWST high-z examples. 

To investigate the nature of the MIRI-selected Si absorption galaxy candidates, we performed SED analysis using the modules listed in Table \ref{t4} with parameter settings used for the AGN candidates (described in section \ref{S:agns}). The observed silicate absorption in 3A galaxies has been associated with differences in the distribution and geometry of the obscuring nuclear dust \citep{Spoon2007}. Spectroscopic studies of low-$z$ samples suggested that while modest silicate absorption can be produced in the clumpy medium of AGN torus, extremely deep silicate absorption can only be produced by a bright source buried in a smooth distribution of geometrically and optically thick dust within the host galaxy \citep{Imanishi2007,Levenson2007,Goulding2012}.  The SEDs of the most deeply buried nuclei can be fitted with radiative transfer models for both the dust surrounding the nuclear source (e.g., the AGN torus) and the additional contributions from the host \citep[e.g.,][and references therein]{GonzalesMartin2019,Efstathiou2022}. 
A broad and deep absorption feature specifically indicates that cold silicate dust with a high optical depth and dust coverage absorbs the bright infrared hot-dust continuum from the AGN at 9.7 $\mu$m \citep[e.g.,][]{Levenson2007, Stalevski2011, GonzalesMartin2013, GonzalesMartin2019}. Therefore, in our SED analysis with CIGALE, the most important component that allows us to decide whether these galaxies have silicate absorption features originating from obscured nuclei is the SKIRTOR AGN \citep{Stalevski2012, Stalevski2016} that is widely used to model AGN MIR SEDs \citep[e.g.,][]{GonzalesMartin2019}. The SKIRTOR model assumes a high-density clumpy and low-density continuous and/or smooth toroidal dust distribution with a mixture of silicate and graphite grains and a standard interstellar medium (ISM) dust composition. The main parameters of the SKIRTOR are the torus inclination ($i$, where $i = 0^{\circ}$ represents a face-on type 1 AGN and $i = 90^{\circ}$ an edge-on type 2 AGN), the torus average edge-on (equatorial) optical depth at 9.7 $\mu$m ($\tau_\mathrm{AGN}$), the torus ratio of outer to inner radius ($ R=R_{\rm{out}}/R_{\rm{in}}$), and the torus half-opening angle ($oa_{\rm{AGN}}$). In this model, the shape of the silicate feature is a function of $i$, $\tau_\mathrm{AGN}$, dust distribution parameters, clump size, and arrangement \citep{Stalevski2011}. If the torus is viewed edge-on, the light from the AGN is absorbed more, resulting in a deeper feature. If viewed face-on, the feature might be shallower due to less dust along the line of sight. A high optical depth means a large fraction of the incident light at 9.7 $\mu$m is being absorbed. For the $R$ parameter the inner radius is assumed to be the dust sublimation radius, a smaller value means a compact torus and a larger value means an extended torus. 
The torus half-opening angle is a measure of the dust-filled region from the equator to the edge of the torus. 

\begin{figure*}
\begin{center}$
\begin{array}{ll}
\includegraphics[scale=0.6]{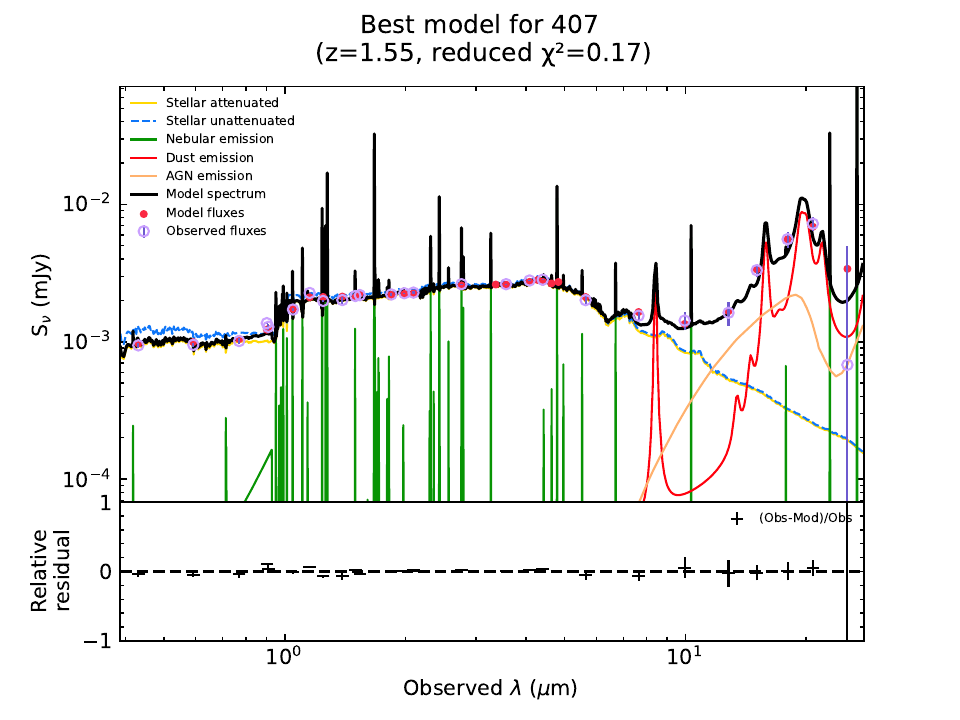}&
\includegraphics[scale=0.6]{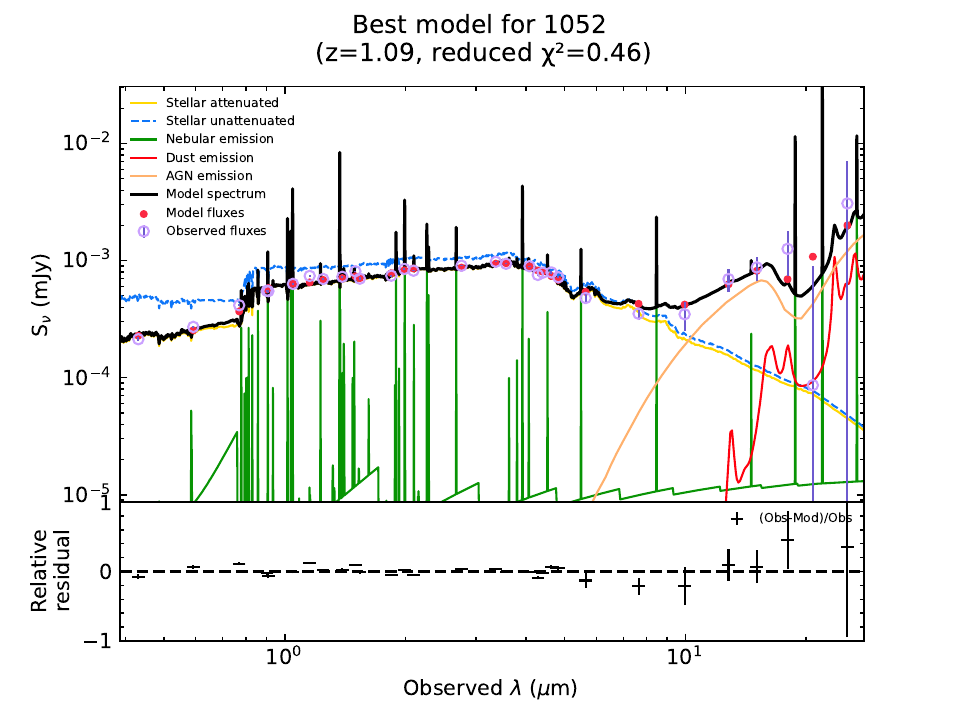}\\
\end{array}$
\end{center}
\caption{ Best-fitting SEDs of the MIRI-selected candidates likely to be Si absorption-dominated galaxies according to SED model parameters.}
    \label{fig:fig6}
\end{figure*}

Based on the measured SED parameters, an edge-on view with $i = 90^{\circ}$ and a high $\tau_\mathrm{AGN}=11$ value identifies the candidate as a heavily obscured AGN with deep silicate absorption. Two galaxies, SMILES IDs 407 and 1052 have $\tau_\mathrm{AGN}=11$, $i = 90^{\circ}$ and $frac_\mathrm{AGN}=0.4 - 0.5$. Based on these parameters, we find it very likely that two (with SMILES IDs 407 and 1052) of the six candidates are Si absorption-dominated galaxies. The obtained best-fitting SEDs of two candidates are shown in Fig. \ref{fig:fig6} support an obscured AGN. 
As seen in these SEDs, the SKIRTOR AGN model with the largest possible $\tau_\mathrm{AGN}$ value of 11 is not extreme enough to produce the very deep silicate absorption feature for these galaxies, because the observed flux is even lower than the absorption feature of the model. \citet[][]{GonzalesMartin2019b} compared SED fitting results of six AGN torus models for a sample of 110 AGN with \textit{Spitzer}/IRS spectroscopy and reported that the SKIRTOR model is among the other five AGN torus models that are not enough to describe the very deep silicate absorption feature. \citet[][]{Efstathiou2022} present SED fitting comparison of different AGN torus models including the CYprus models for Galaxies and their NUclear Spectra (CYGNUS) and report that most of the AGN torus models such as SKIRTOR do not produce deep enough silicate absorption features for their ULIRG sample. 
Identifying and eliminating the limits of current AGN torus models is beyond the scope of this work. Among the existing models, the SKIRTOR model is generally sufficient to provide a general understanding of the nature of the identified candidates, which is our goal.
However, it is not possible to reach a definitive conclusion about the silicate strength of our candidates without MIR spectroscopy. However, these galaxies have a very high probability of having a deeper silicate absorption feature than the deepest feature that can be obtained from the SKIRTOR model; therefore we interpret these sources as strong Si absorption-dominated galaxy candidates. 

Here, we have demonstrated the power of MIRI colors to find Si absorption-dominated galaxies beyond the local Universe. 
\citet[][]{GarciaBernete2022} presents JWST MIRI color diagrams to select the most obscured galaxy nuclei in CONs based on CYGNUS SED model fitting results of local U/LIRGs selected from the IDEOS sample. Their color limits are model dependent while our color measurements are derived from observed MIR spectra.
We also note that their sample was selected to be galaxies with high total IR luminosities.
Since the deep Si absorption is also expected to be a prominent feature in the spectra of CONs in local U/LIRGs, our $z=1$ $\log(\rm{F}(21\,\mu$m)$/\rm{F}(15\,\mu$m)) < -0.7 color limit agrees well with their CONs limit of $\log(\rm{F}(21\,\mu$m)$/\rm{F}(15\,\mu$m)) < -1.0 \citep[][]{GarciaBernete2022}. \citet[][]{GarciaBernete2024} present  JWST color--color diagrams for selecting deeply obscured nuclei at higher redshifts, however we can not make a direct comparison because of the different color bands used.
We note that these galaxies might also be CONs, that have supermassive black hole accretion and/or compact star-forming activity which is completely obscured by dust. MIRI spectroscopy is needed to apply CONs criteria \citep[e.g.,][and references therein]{Donnan2023} to find out if these are completely obscured nuclei. 

As a result of the SED analysis, we found that the other three Si absorption-dominated galaxy candidates have no AGN contribution, their $frac_\mathrm{AGN}=0.0$. These galaxies might be dusty star-forming galaxies with a deep silicate absorption feature since the silicate absorption might also be produced by gas and dust in the host galaxy \citep[e.g.,][]{Levenson2007, Goulding2012}. These galaxies (SMILES IDs 1141, 1323, 2677) might contain buried nuclei of strong star-forming processes.

\subsection{Comparison of MIRI color selection results with emission line ratio AGN and/or galaxy diagnostics}\label{S:comp}

To compare our MIRI color selection with other AGN and galaxy classification methods based on emission line ratios we used the available [O\,{\sc iii}]$\lambda5007$, $H\beta$,  emission line fluxes of 13 AGN (four at $z=0.28$ and nine at $z=0.62$) and ten SFGs ( one at $z=0.28$ and nine at $z=0.62$) from MUSE Hubble Ultra Deep Field surveys DR 2 Main source catalog. We compared our AGN and SFG selection with that of \citet{Juneau2014}. For $z=0.28$, we used their low-redshift criteria (their figure 1) and for $z=0.62$  we used their intermediate-z criteria (their figure 4). All of our SFGs candidates are located in their SFGs region, so our SFG selection is consistent with their method. The agreement is not good for AGN, only 3/13 of our AGN are lying in their AGN region (both are 1A). However, they also report more discrepancies with IR-detected AGN; only 1/6 in the AGN region. Our ratio of IR-detected galaxies lying in the AGN region is close to theirs. As stated by \citet{Juneau2014}, some IR AGN candidates may have contamination by SFGs; however, we note that our candidates' existing spectral data are insufficient to reach a full conclusion.
Based on the limited number of spectra, this comparison shows a possibility of misclassifications at least for some of the AGN candidates.

\citet{Mazzolari2024} presented a new AGN selection technique based on [O\,{\sc iii}]$\lambda4363$ which is accessible with JWST. Unfortunately, none of our MIRI selected AGN and SFG candidates have NIRSpec measurements in JADES Data Release 3 (DR3) catalog. Therefore, we can not make a direct comparison with the [O\,{\sc iii}]$\lambda4363$/$H\gamma$ ratio classification \citep{Mazzolari2024}. To further expand its place in the broader context of AGN selection techniques, the MIR-based method presented here will be compared in detail with the optical-based methods \citep[e.g.,][]{Juneau2014,Mazzolari2024} when optical spectra become available in the future.

\subsection{Physical interpretation of MIR colors in the framework of galaxy evolution}\label{S:physical}
% I AM HERE
MIR wavelengths (~3–30 µm) are sensitive to dust properties, star formation activity, AGN heating. Therefore, the MIR color classification of galaxies reveal evolutionary stages through distinct physical mechanisms: strong PAH emission indicates star formation, silicate absorption and dust geometry mark obscured transitional phases, and suppressed PAH features with elevated continua signal AGN emergence \citet[e.g.,][]{Spoon2006,Spoon2022}. In this work, we followed \citet[][]{Spoon2022} and investigated new MIRI colors based on the relative strengths of Si absorption feature and PAH features that provide insights into the interstellar medium (ISM) conditions and energy sources in galaxies. The six colors we show in Fig. \ref{fig:fig1} and Fig. \ref{fig:newfig2}, traces Si97 at different redshifts. In each panel the color in the y-axis traces the Si strength; the more negative colors means higher Si strength. Therefore, these colors reveal the obscuration levels at different types of galaxies, the most obscured galaxies clearly found closer to the bottom region. This is also relevant for the colors on the x-axis; they trace the Si strength in the opposite direction, toward the left: the more negative value means higher Si strength. Therefore, the obscuration level increases toward the left in these panels. In most panels AGN distribute around color values of 0 as expected; they exhibit stronger MIR continuum emission due to dust heated by the AGN activity and exhibit weaker/non PAH features. The MIR colors on the y-axis also trace the continuum-to-PAH ratio; lower values mean PAH emission is higher. In contrast, higher values toward the upper region mean low ratios as expected in AGN. Therefore, SFGs have lower values compared to AGN. In Fig. \ref{fig:fig1} we also investigated the colors of "transition" galaxies (classes 2A–3B) with mixed MIR signatures, suggesting an evolutionary link between starburst-dominated and AGN-dominated systems.
These galaxies may represent a phase where AGN feedback begins to quench star formation, marking a critical stage in galaxy evolution.

\citet[][]{Spoon2022} demonstrated the use of MIR color--color diagrams to classify galaxies based on their dominant energy sources: star formation, AGN activity, or a mix of both. We extended their MIR color diagnostics to intermediate redshifts with recent JWST observations. Our MIRI color diagnostics that can distinguish between AGN (with hot dust continuum emission), SFGs (dominated by PAH features) and, Si absorption dominated galaxies are consistent with the diagnostics of \citet[][]{Spoon2022}. Galaxy-evolution scenarios \citep[e.g.,][]{Hopkins2008a} suggest that when normal SFGs (e.g., 1C class) merge, they become  "transition" galaxies (2A–3B classes) going through strong nuclear obscuration before ending up as a naked AGN (class 1A). So our MIRI color diagrams at intermediate redshifts can be used to identify galaxies at different stages of galaxy evolution. 

\section{Conclusions}\label{S:conc}

We obtained $z$-dependent ($0.25< z <2.10$) \textit{JWST}/MIRI color diagnostics based on \textit{Spitzer} spectral features.
We applied the MIR spectral classification diagnostics obtained in \S \ref{S:colcol} to classify galaxies in the SMILES survey. We have made the classification code and the GMM region files publicly available.\footnote {\url{https://github.com/Ecekilerci/MIRIcolordiagnostics}}
Our conclusions are summarized below. 

\begin{enumerate}

\item We show that combinations of MIRI colors alone can identify different MIR classes when the redshift is known. 

\item Our MIRI color limits provide an effective tool to separate samples of SFGs (1C \& 2C ), AGN (1A \& 1B ), and Si absorption-dominated galaxies (3A) at cosmic noon.

\item Compared to previously known 3A galaxies observed with \textit{Spitzer}, our 3A candidates are the first JWST high-redshift examples to date.

\item The MIRI-selected SFGs identified in this work show that JWST finds a diversity of high-z SFGs, including those with lower masses and IR luminosities. 

\item The majority of MIRI-selected SFGs at $0.9 < z < 1.6$  identified in this work, are consistent with the $z \sim 1$ MS. Our MIRI-selected SFGs at $z \sim 2$ indicate an evolution compared to $z \sim 1$ MS. 

   \end{enumerate}

We used MIRI's unique continuous MIR filter coverage and obtained improved color diagnostics for MIR galaxy classification. As a future prospect, we will apply the $z$-dependent JWST/MIRI color diagnostics introduced in this work to other completed and future JWST/MIRI survey data. We also aim to improve our method with NIRspec emission line measurements to be obtained in the future. Increasing the number of MIRI classified SFGs, AGN, and  Si absorption-dominated galaxies toward cosmic noon will contribute to our understanding of galaxy evolution by revealing the characteristics of larger samples.

\begin{acknowledgements}
We thank the anonymous referee for carefully reading the manuscript and for suggestions and recommendations, which have substantially improved the quality and clarity of the paper.
TG acknowledges the support of the National Science and Technology Council of Taiwan through grants 113-2112-M-007 -006.
TH acknowledges the support of the National Science and Technology Council of Taiwan through grants 110-2112-M-005-013-MY3, 110-2112-M-007-034-, 113-2123-M-001-008-, and 113-2112-M-005-009-MY3.
This work is based on observations made with the NASA/ESA/CSA James Webb Space Telescope. The data were obtained from the Mikulski Archive for Space Telescopes at the Space Telescope Science Institute, which is operated by the Association of Universities for Research in Astronomy, Inc., under NASA contract NAS 5-03127 for JWST. These observations are associated with the program ERO. 
The authors acknowledge the CEERS team for developing their observing program with a zero-exclusive-access period.
This work is based on observations taken by the CANDELS Multi-Cycle Treasury Program with the NASA/ESA \textit{HST}, which is operated by the Association of Universities for Research in Astronomy, Inc., under NASA contract NAS 5-26555.
Some of the data presented in this paper were obtained from the Mikulski Archive for Space Telescopes (MAST) at the Space Telescope Science Institute. The specific observations analyzed can be accessed via doi:10.17909/agda-2w34.
\end{acknowledgements}

\bibliography{obsagn}{}
\bibliographystyle{aa}

\begin{appendix} 
\onecolumn
\section{PAH equivalent width correlations}\label{S:appendix}

\begin{figure*}[h!]
\begin{center}$
\begin{array}{lll}
\includegraphics[scale=0.65]{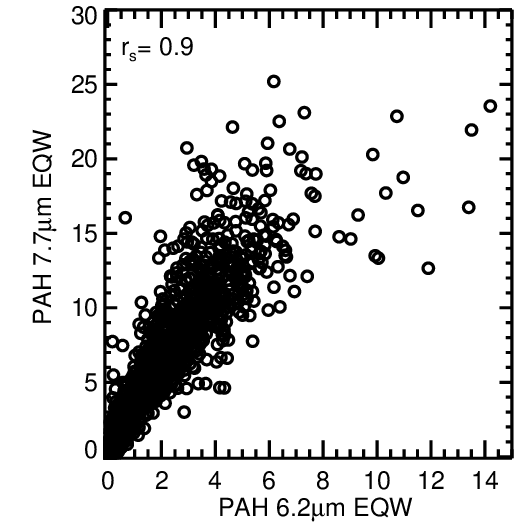}&
\includegraphics[scale=0.65]{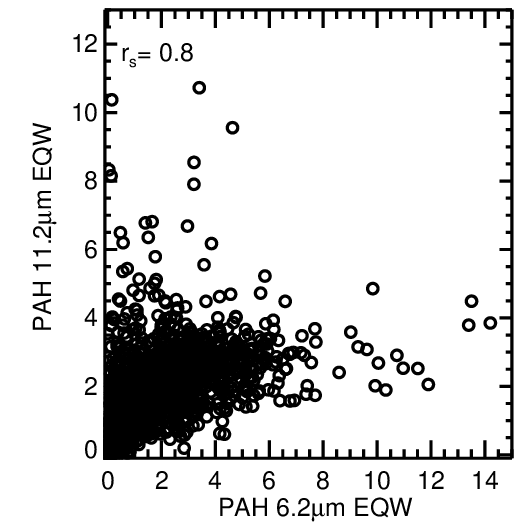}&
\includegraphics[scale=0.65]{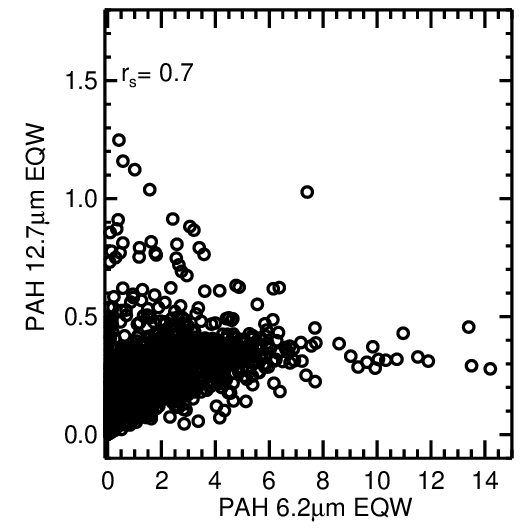}\\
\end{array}$
\end{center}
\caption{ Correlations between PAH62 EQW and PAH77 EQW (left panel), PAH62 EQW and PAH112 EQW (middle panel), PAH62 EQW and PAH127 EQW (right panel). The Spearman’s rank correlation coefficient r$_{s}$ is listed in each panel. The significance of r$_{s}$ is 0.00 for each panel. }
\label{fig:A1}
\end{figure*}

As shown in Fig. \ref{fig:A1} PAH62 EQW positively correlates with PAH77 EQW (left panel), PAH112 EQW (middel panel), and PAH127 EQW (right panel). The strength of the correlations is given by Spearman’s rank correlation coefficient (r$_{s}$) in each panel. The r$_{s} \geq 0.7$ values indicate a strong correlation between the EQW measurements.
A significance value of 0.00 is found for each panel. Therefore, there is less than a 0.1\% chance that the found r$_{s}$ happened by chance if there were no correlations between these EQW measurements.

\FloatBarrier
\twocolumn
%____________________________________________________________
% A long table in appendix
%------------------------------------------------------------
% This is the start of the page
\onecolumn
\section{Gaussian mixture model regions}\label{S:gmmfit}

The color--color diagrams in Fig. \ref{fig:fig1} and Fig. \ref{fig:newfig2} trace Si97 and one of the PAH62, PAH77, PAH112, PAH127 emission features for the given redshift range. 
GMM model regions of each MIR class in MIRI color--color diagrams shown in Fig. \ref{fig:fig1} are listed in Table \ref{t1}.  GMM model regions of AGN  (1A \& 1B \& 2A) and SFGs (1C \& 2C) shown in Fig. \ref{fig:newfig2} are given in Table \ref{t2}.

GMM model regions of AGN  (1A \& 1B \& 2A) and SFGs (1C \& 2C) on the color--color diagrams adopted from \citet[][]{Kirkpatrick2017b} and \citet[][]{Kirkpatrick2023} are listed in Table \ref{t3}.

\begin{longtable}{p{3.4cm}p{4.4cm}rrllr}
\caption{\label{t1}MIRI color regions of the distinct regions for 1A, 1B, 1C, 2A, 2B, 2C, 3A, 3B MIR spectral classes obtained by GMM analysis.}\\
MIR Spectral Class  & GMM Component Number     & Center-$x$ & Center-$y$ & Width & Height & Angle\\ 
\hline 
\endfirsthead
\caption{continued.}\\
\hline
MIR spectral class  & GMM component number     & Center-$x$ & Center-$y$ & Width & Height & Angle\\[1ex] 
\hline
\endhead
\hline
\endfoot
\hline
\hline
$z=0.25-0.30$ & $x=\log(\rm{F}(12\,\mu$m)$/\rm{F}(07\,\mu$m)) & $y=\log(\rm{F}(12\,\mu$m)$/\rm{F}(10\,\mu$m)) & & & & \\
\hline
1A & 1& 0.2136 & 0.1010 & 0.7725 & 0.2554 & -151.76\\
1B &1 &0.1405& -0.2714 &0.7421 &0.2193 &51.14\\
1B&2 &0.2956& 0.0346 &0.8799 &0.2484 &-148.81\\
1C&1 &0.0668& -0.4617 &0.6951 &0.1042 &46.42\\
2A&1 &-0.1108& -0.3437 &1.1263 &0.2648 &-137.45\\
2B&1 &-0.1437& -0.5728 &0.9194 &0.2671 &45.63\\
2C&1 &-0.1486& -0.6766 &0.6242 &0.0814 &46.96\\
3A&1 &-1.0145& -1.3882 &1.1161 &0.1251 &-159.60\\
3A&2 &-0.6624& -1.0225 &0.8311 &0.4586 &-148.22\\
3B&1 &-0.5175& -0.9988 &0.6665 &0.2075 &46.41\\
\hline 
\hline
$z=0.58-0.65$ & $x=\log(\rm{F}(15\,\mu$m)$/\rm{F}(10\,\mu$m)) & $y=\log(\rm{F}(15\,\mu$m)$/\rm{F}(12\,\mu$m)) & & & & \\
\hline
1A &1 &0.2223& 0.0901 &0.4058 &0.1626 &-152.83\\
1A &2 &-0.0362& -0.0349 &0.4776 &0.2875 &168.31\\
1B &1 &0.0572& -0.2706 &0.5917 &0.1821 &48.39\\
1B &2 &0.2198& -0.0126 &0.7063 &0.2175 &-151.94\\
1C &1 &-0.0171& -0.4208 &0.5500 &0.0925 &-136.46\\
2A &1 &-0.1363& -0.2969 &0.8477 &0.2094 &-138.03\\
2B &1 &-0.1623& -0.4960 &0.7192 &0.2250 &-136.37\\
2C &1 &-0.1790& -0.5753 &0.4598 &0.0780 &-135.77\\
3A &1 &-0.5546& -0.8427 &0.6569 &0.4289 &54.90\\
3A &2 &-0.9630& -1.0037 &0.7690 &0.1308 &-135.06\\
3B &1 &-0.4387& -0.8116 &0.5855 &0.1887 &-142.92\\
\hline 
\hline
$z=0.58-0.65$ & $x=\log(\rm{F}(15\,\mu$m)$/\rm{F}(10\,\mu$m)) & $y=\log(\rm{F}(15\,\mu$m)$/\rm{F}(21\,\mu$m)) & & & & \\
\hline
1A &1 &0.1645& -0.1706 &0.7842 &0.4558 &125.10\\
1B &1 &0.1316& -0.3287 &0.6184 &0.5290 &174.19\\
1C &1 &-0.0175& -0.3730 &0.4089 &0.3226 &-163.88\\
2A &1 &-0.2511& -0.6972 &0.5584 &0.4719 &160.99\\
2A &2 &-0.0446& -0.5329 &0.4604 &0.4141 &148.56\\
2A &3 &0.0216& 0.4006 &0.0040 &0.0040 &90.00\\
2B &1 &-0.1595& -0.6949 &0.5957 &0.4359 &-137.49\\
2C &1 &-0.1790& -0.6005 &0.3497 &0.2768 &-150.99\\
3A &1 &-0.5385& -1.1491 &0.6612 &0.4583 &62.40\\
3A &2 &-0.8235& -0.8866 &0.8627 &0.3868 &-158.73\\
3B &1 &-0.4387& -1.0327 &0.5965 &0.4613 &69.04\\
\hline 
\hline
$z=0.90-0.92$ & $x=\log(\rm{F}(18\,\mu$m)$/\rm{F}(15\,\mu$m)) & $y=\log(\rm{F}(18\,\mu$m)$/\rm{F}(25\,\mu$m)) & & & & \\
\hline
1A &1 &0.0620& -0.1843 &0.7919 &0.3452 &84.79\\
1B &1 &-0.1690& -0.3658 &0.7754 &0.5570 &-146.37\\
1C &1 &-0.4600& -0.4370 &0.4368 &0.3290 &-152.04\\
2A &1 &-0.4343& -0.7607 &0.6632 &0.3858 &-151.14\\
2A &2 &-0.2362& -0.5559 &0.4416 &0.3182 &80.96\\
2A &3 &-0.2645& 0.7378 &0.0040 &0.0040 &90.00\\
2B &1 &-0.4399& -0.6913 &0.4711 &0.4005 &-166.93\\
2B &2 &-0.6278& -0.8295 &0.5723 &0.3141 &62.65\\
2C &1 &-0.6311& -0.6842 &0.3964 &0.2746 &-147.89\\
3A &1 &-1.0050& -1.2608 &0.9251 &0.0385 &78.03\\
3A &2 &-0.9489& -1.1386 &0.9224 &0.6395 &-150.03\\
3B &1 &-0.8938& -1.1434 &0.6249 &0.3745 &64.89\\
\hline 
\hline
$z=1.0-1.10$ & $x=\log(\rm{F}(21\,\mu$m)$/\rm{F}(12\,\mu$m)) & $y=\log(\rm{F}(21\,\mu$m)$/\rm{F}(15\,\mu$m)) & & & & \\
\hline
1A &1 &0.2755& 0.1876 &0.5124 &0.1308 &-143.85\\
1A &2 &-0.0183& -0.0180 &0.5535 &0.2056 &-164.56\\
1B &1 &0.1921& -0.0708 &0.6050 &0.1593 &49.85\\
1B &2 &0.3074& 0.1401 &0.8460 &0.1588 &-145.31\\
1C &1 &0.1106& -0.2261 &0.3317 &0.0751 &-135.35\\
1C &2 &0.1787& -0.1548 &0.8551 &0.0917 &46.77\\
2A &1 &-0.0363& -0.1863 &0.3457 &0.1764 &49.30\\
2A &2 &-0.3299& -0.4754 &0.5355 &0.2070 &51.26\\
2A &3 &0.1293& -0.0443 &0.4671 &0.0651 &-136.98\\
2B &1 &0.0105& -0.2834 &0.6071 &0.1680 &48.29\\
2B &2 &-0.1679& -0.4804 &0.6337 &0.1437 &-145.14\\
2C &1 &-0.0439& -0.3904 &0.4944 &0.0639 &47.38\\
3A &1 &-0.4238& -0.7611 &0.4917 &0.3458 &-144.12\\
3A &2 &-0.8580& -1.0474 &0.7736 &0.2880 &46.86\\
3B &1 &-0.3500& -0.7115 &0.6622 &0.1509 &-137.49\\
\hline 
\hline
$z=1.5-1.6$ & $x=\log(\rm{F}(25\,\mu$m)$/\rm{F}(15\,\mu$m)) & $y=\log(\rm{F}(25\,\mu$m)$/\rm{F}(21\,\mu$m)) & & & & \\
\hline
1A &1 &0.2665& 0.1009 &0.5124 &0.2261 &-155.38\\
1A &2 &-0.1073& -0.0384 &0.4313 &0.2235 &140.32\\
1B &1 &0.1664& -0.2492 &0.6474 &0.2368 &49.84\\
1B &2 &0.2948& 0.0175 &0.8338 &0.2421 &-155.00\\
1C &1 &0.0879& -0.4460 &0.4210 &0.0955 &-135.77\\
1C &2 &0.2267& -0.2888 &0.9494 &0.1558 &-137.10\\
2A &1 &-0.0523& -0.2636 &0.3369 &0.2654 &-144.10\\
2A &2 &-0.3449& -0.5321 &0.6425 &0.3545 &45.07\\
2A &3 &0.1171& -0.1797 &0.5414 &0.1435 &-136.15\\
2B &1 &-0.0024& -0.4005 &0.6143 &0.2960 &-140.51\\
2B &2 &-0.1811& -0.6134 &0.6439 &0.2318 &-148.39\\
2C &1 &-0.0935& -0.6097 &0.5508 &0.0831 &45.24\\
3A &1 &-0.8536& -1.1305 &1.0549 &0.4287 &-150.87\\
3A &2 &-0.5237& -0.8977 &0.6504 &0.4879 &-143.91\\
3B &1 &-0.4423& -0.9013 &0.5981 &0.2219 &-136.61\\
\hline
\end{longtable}

% TABLE B.2 
\begin{longtable}{p{3.4cm}p{4.4cm}rrllr}
\caption{\label{t2}MIRI color regions of AGN  (1A \& 1B \& 2A) and SFGs (1C \& 2C) obtained by GMM analysis.}\\
MIR Spectral Class  & GMM Component Number     & Center-$x$ & Center-$y$ & Width & Height & Angle\\ 
\hline 
\endfirsthead
\caption{continued.}\\
\hline
MIR spectral class  & GMM component number     & Center-$x$ & Center-$y$ & Width & Height & Angle\\[1ex] 
\hline
\endhead
\hline
\endfoot
\hline
\hline
$z=0.25-0.30$ & $x=\log(\rm{F}(12\,\mu$m)$/\rm{F}(07\,\mu$m)) & $y=\log(\rm{F}(12\,\mu$m)$/\rm{F}(10\,\mu$m)) & & & & \\
\hline
AGN &1 &0.0636& -0.1598 &0.8716 &0.5999 &46.50\\
AGN &2 &0.2799& 0.1273 &0.5936 &0.2344 &-151.67\\
SFG &1 &0.0284& -0.5036 &0.6164 &0.0925 &45.43\\
SFG &2 &0.1130& -0.3914 &1.4444 &0.1360 &45.93\\
\hline 
\hline
$z=0.58-0.65$ & $x=\log(\rm{F}(15\,\mu$m)$/\rm{F}(10\,\mu$m)) & $y=\log(\rm{F}(15\,\mu$m)$/\rm{F}(12\,\mu$m)) & & & & \\
\hline
AGN &1 &-0.1287& 0.0098 &0.1682 &0.0899 &-165.61\\
AGN &2 &0.2039& 0.0867 &0.3139 &0.1347 &-151.38\\
AGN &3 &0.1298& -0.1029 &0.9882 &0.3110 &-135.05\\
SFG &1 &-0.0510& -0.4543 &0.4295 &0.0828 &-137.31\\
SFG &2 &0.0187& -0.3816 &1.0395 &0.1168 &-136.05\\
\hline 
\hline
$z=0.58-0.65$ & $x=\log(\rm{F}(15\,\mu$m)$/\rm{F}(10\,\mu$m)) & $y=\log(\rm{F}(15\,\mu$m)$/\rm{F}(21\,\mu$m)) & & & & \\
\hline
AGN &1 &0.1638& -0.1813 &0.7936 &0.4391 &127.87\\
AGN &2 &0.1474& -99.0000 &0.5178 &0.0040 &0.00\\
AGN &3 &0.0774& -0.3832 &0.8953 &0.4909 &46.58\\
SFG &1 &-0.0381& -0.4021 &0.5320 &0.3431 &-135.69\\
\hline 
\hline
$z=0.90-0.92$ & $x=\log(\rm{F}(18\,\mu$m)$/\rm{F}(15\,\mu$m)) & $y=\log(\rm{F}(18\,\mu$m)$/\rm{F}(25\,\mu$m)) & & & & \\
\hline
AGN &1 &0.0711& -0.1735 &0.7970 &0.2963 &91.06\\
AGN &2 &-0.2021& -0.4180 &0.8312 &0.5029 &-137.04\\
SFG &1 &-0.4818& -0.4686 &0.5795 &0.3402 &46.54\\
\hline 
\hline
$z=1.0-1.10$ & $x=\log(\rm{F}(21\,\mu$m)$/\rm{F}(12\,\mu$m)) & $y=\log(\rm{F}(21\,\mu$m)$/\rm{F}(15\,\mu$m)) & & & & \\
\hline
AGN &1 &0.2195& 0.0292 &1.0563 &0.2441 &45.31\\
AGN &2 &0.2509& 0.1749 &0.4313 &0.1061 &-142.75\\
AGN &3 &-0.1217& -0.0240 &0.2242 &0.0531 &-148.39\\
SFG &1 &0.1263& -0.2096 &0.5726 &0.0836 &46.01\\
SFG &2 &0.0100& -0.3338 &0.5926 &0.0429 &47.60\\
\hline 
\hline
$z=1.5-1.6$ & $x=\log(\rm{F}(25\,\mu$m)$/\rm{F}(15\,\mu$m)) & $y=\log(\rm{F}(25\,\mu$m)$/\rm{F}(21\,\mu$m)) & & & & \\
\hline
AGN &1 &0.0318& 0.0068 &0.7793 &0.1318 &-177.57\\
AGN &2 &0.2612& 0.1185 &0.3348 &0.1401 &-160.24\\
AGN &3 &0.2078& -0.1020 &1.1188 &0.3863 &-136.29\\
SFG &1 &0.0736& -0.4591 &0.5359 &0.0920 &-136.42\\
SFG &2 &0.1315& -0.3798 &1.2232 &0.1519 &-135.83\\
\hline
\end{longtable}

% TABLE B.3 
\begin{longtable}{p{3.4cm}p{4.4cm}rrllr}
\caption{\label{t3}MIRI color regions of AGN  (1A \& 1B \& 2A) and SFGs (1C \& 2C) on the color diagrams adopted from the literature  obtained by GMM analysis.}\\
MIR Spectral Class  & GMM Component Number     & Center-$x$ & Center-$y$ & Width & Height & Angle\\ 
\hline 
\endfirsthead
\caption{continued.}\\
\hline
MIR spectral class  & GMM component number     & Center-$x$ & Center-$y$ & Width & Height & Angle\\[1ex] 
\hline
\endhead
\hline
\endfoot
\hline
\hline
$z=1.0-1.1$ & $x=\log(\rm{F}(15\,\mu$m)$/\rm{F}(07\,\mu$m)) & $y=\log(\rm{F}(18\,\mu$m)$/\rm{F}(10\,\mu$m)) & & & & \\
\hline
AGN &1 &0.3160& 0.2809 &1.5998 &0.5815 &-166.66\\
SFG &1 &0.1559& 0.6351 &1.9786 &0.8027 &-158.90\\
\hline 
\hline
$z=1.4-1.5$ & $x=\log(\rm{F}(12\,\mu$m)$/\rm{F}(10\,\mu$m)) & $y=\log(\rm{F}(18\,\mu$m)$/\rm{F}(10\,\mu$m)) & & & & \\
\hline
AGN &1 &0.1279& 0.3031 &0.6796 &0.0968 &65.77\\
AGN &2 &0.1363& 0.4754 &3.2915 &0.6395 &47.97\\
SFG &1 &-0.6842& 0.0817 &1.7547 &0.4047 &56.38\\
SFG &2 &-0.3076& 0.6029 &3.4785 &0.2054 &-135.91\\
\hline 
\hline
$z=2.0-2.1$ & $x=\log(\rm{F}(18\,\mu$m)$/\rm{F}(10\,\mu$m)) & $y=\log(\rm{F}(21\,\mu$m)$/\rm{F}(15\,\mu$m)) & & & & \\
\hline
AGN &1 &0.1827& 0.1853 &2.0844 &0.6643 &-176.12\\
SFG &1 &-0.4165& 0.7037 &2.0901 &0.8823 &-164.58\\
\hline
\end{longtable}

\end{appendix}
\end{document}